\documentclass[final,5p,times,twocolumn]{elsarticle}
\usepackage{amsmath,amsfonts}
\usepackage{algorithmic}
\usepackage{array}
\usepackage[caption=false,font=footnotesize,labelfont=sf,textfont=sf]{subfig}
\usepackage{graphicx}
\usepackage{textcomp}
\usepackage{url}
\usepackage{verbatim}
\usepackage{tabularx} 
\usepackage{hyperref}
\def\BibTeX{{\rm B\kern-.05em{\sc i\kern-.025em b}\kern-.08em
    T\kern-.1667em\lower.7ex\hbox{E}\kern-.125emX}}
\usepackage{balance}

\usepackage{amsmath}	
\usepackage{amssymb}	
\usepackage{caption}
\usepackage{color}     

\usepackage{gensymb}
\usepackage{graphicx}	
\usepackage{multirow}
\usepackage{dblfloatfix}

\usepackage[numbers]{natbib}  
\usepackage{svg}

\PassOptionsToPackage{dvipsnames,table}{xcolor}
\usepackage{float} 
\usepackage{xcolor} 
\usepackage{placeins} 
\usepackage{cite}
\renewcommand{\vec}[1]{\mathbf{#1}}
\newcommand{\hvec}[1]{\hat{\mathbf{#1}}}
\newcommand{\DS}{\displaystyle}

\newcommand{\HALF}{\frac{1}{2}}

\newcommand{\av}[1]{\left< {#1} \right>}
\newcommand{\quotes}[1]{``#1''}
\newcommand{\tens}[1]{\mathsf{#1}}

\newcommand{\cE}{\mathcal{E}}

\newcommand{\cL}{\mathcal{L}}

\newcommand{\cR}{\mathcal{R}}
\newcommand{\cS}{\mathcal{S}}

\newcommand{\cU}{\mathcal{U}}
\newcommand{\cV}{\mathcal{V}}

\newcommand{\cc}{ {\boldsymbol{c}} }

\newcommand{\gpluto}{gPLUTO}
\newcommand{\pluto}{PLUTO}

\usepackage{makecell}
\usepackage{colortbl}

\definecolor{redcell}{rgb}{0.96,0.6,0.6}
\definecolor{orangecell}{rgb}{1.0,0.8,0.4}
\definecolor{yellowcell}{rgb}{1.0,1.0,0.6}
\definecolor{greencell}{rgb}{0.7,1.0,0.7}

\usepackage{booktabs} 
\usepackage{pgfplots}
\usepackage{listings}

\newcommand{\Cinline}[1]{\lstinline[language=C, basicstyle=\ttfamily\small]{#1}}

\lstnewenvironment{Ccode}
{\lstset{
  language=C,
  emph={double},
  emphstyle={\color{black}},
  keywordstyle=[2]{\color{red}},
  keywords=[2]{pragma,acc} 
  directivestyle=\color{brown},
  commentstyle=\color{blue},
  basicstyle=\ttfamily\small,
   tabsize=2
}} {}

\usepackage{listings}
\usepackage{xcolor}

\lstdefinestyle{gplutostyle}{
    language=C++,
    basicstyle=\ttfamily\footnotesize, 
    columns=flexible,                  
    keepspaces=true,                   
    numbers=none,                      
    frame=single,                      
    rulecolor=\color{black},
    keywordstyle=\color{green!60!black}, 
    commentstyle=\color{gray},
    stringstyle=\color{purple},
    moredelim=*[l][\color{red}]{\#},
    showstringspaces=false
}

\lstnewenvironment{ccode}[1][]
{\lstset{style=gplutostyle, #1}}
{}

\newcounter{bla}

\usepackage{amssymb}
\usepackage{comment}
\makeatletter
\def\ps@pprintTitle{%
   \baselineskip=12pt
   \textnormal{}%
}
\makeatother

\pgfplotsset{
    extra xlabel near ticks/.style={
        extra x tick style={
            tick label style={
                yshift=0pt 
            }
        }
    }
}
\pgfplotsset{compat=1.18}

\begin{document}

\begin{frontmatter}

\title{The PLUTO Code on GPUs: A First Look at Eulerian MHD Methods}
\author{M. Rossazza$^{1}$,
A. Mignone$^{1}$,
M. Bugli$^{2,1,3}$,
S. Truzzi$^{1}$,
L. Riha$^{4}$,
T. Panoc$^{4}$,
O. Vysocky$^{4}$,
N. Shukla$^{5}$,
A. Romeo$^{5}$,
V. Berta$^{1}$
\\
$^{1}$Dipartimento di Fisica, Università degli Studi di Torino, Via Pietro Giuria 1, I-10125 Torino, Italy\\
$^{2}$Institut d’Astrophysique de Paris, UMR 7095, CNRS \& Sorbonne Universit\'e, 98 bis bd Arago, 75014 Paris, France\\
$^{3}$INFN - sezione di Torino, Via Pietro Giuria 1, I-10125 Torino, Italy\\
$^{4}$IT4Innovations, VSB - Technical University of Ostrava, Ostrava, Czech Republic\\
$^{5}$CINECA, Via Magnanelli 6/3, 40033 Casalecchio di Reno, BO, Italy\\
Corresponding author: Marco Rossazza, marco.rossazza@unito.it
}


\begin{abstract}
We present preliminary performance results of \gpluto{}, the new GPU-optimized implementation of the \pluto{} code for computational plasma astrophysics.
Like its predecessor, \gpluto{} employs a finite-volume formulation to numerically solve the equations of magnetohydrodynamics (MHD) in multiple spatial dimensions.
Still, this new implementation is a complete rewrite in C++ and leverages the OpenACC programming model to achieve acceleration on NVIDIA GPUs.
While a more comprehensive description of the code and its several other modules will be presented in a future paper, here we focus on some preparatory results that demonstrate the code potential and performance on pre exa-scale parallel architectures.
\end{abstract}

\end{frontmatter}

\section{Introduction}
%
%

Magnetohydrodynamics (MHD) simulations play a pivotal role in understanding a vast range of complex physical phenomena. 
In the context of astrophysical processes, these include the formation and evolution of planets~\citep{lesur2021}, stars~\citep{teyssier2019}, and galaxies~\citep{zhang2018}, the acceleration of high-energy particles~\citep{lazarian2012,amato2018}, and the multi-wavelength emission associated to the activity of compact objects such as black holes~\citep{narayan2005} and neutron stars~\citep{faber2012,uzdensky2014}. 
There are also numerous terrestrial applications that exploit the predictive power of the MHD framework, such as nuclear fusion~\citep{knaepen2008,inghirami2016}, space weather predictions~\citep{shibata2011}, and plasma confinement~\citep{doyle2007}. 

The computational demands of MHD simulations are inherently tied to the need to describe at once spatial and temporal scales that often span many orders of magnitude.
The ability to resolve phenomena occurring at scales much smaller than the typical size of the global system (e.g., the turbulent structures in the plasma constituting an accretion disk or a jet propagating through the interstellar medium) plays a crucial role in enabling a far deeper understanding of their properties and allowing more accurate predictions of their observational/experimental signatures.  
Prime examples in the past decade of the qualitative impact of higher computational efficiency in multi-dimensional astrophysical MHD simulations include models of stellar explosions~\citep{mezzacappa2023,janka2025}, dynamo processes~\citep{charbonneau2020,delzanna2022}, and accretion/ejection of matter around central compact objects~\citep{mayer2019,cattorini2024}.      
In recent years, the field of high-performance computing has witnessed a transformative shift thanks to the advent of Graphics Processing Units (GPUs) that offer massive parallelism and computational capabilities. 
The use of GPUs in scientific computing has indeed opened up new frontiers in numerical simulations, enabling researchers to explore complex physical systems at unprecedented resolutions and speeds.
Notable examples are the codes H-AMR~\citep{liska2022}, IDEFIX~\citep{Lesur_2023}, AthenaK~\citep{stone2024}, ECHO~\citep{delzanna2024}, and AsterX~\citep{kalinani2025}.

In this paper, we present the latest noteworthy advancements in the development of the GPU-ready version of the \pluto{} code~\citep{Mignone_PLUTO2007,Mignone_PLUTO2012} for modeling astrophysical plasmas.  
\pluto{} has been widely used by hundreds of researchers in the astrophysical community over the past two decades.
Owning to the variety of physical modules provided (i.e., from classical to relativistic flows, ideal and non-ideal dynamics, different geometries, and equations of state), the code has been employed in very different astrophysical contexts, ranging from high-energy astrophysics~\citep{2010_Mignone,Mignone_2013,Mattia_2023,Pavan_2023} to young stellar objects~\citep{orlando2011}, supernova remnants~\citep{2018_Olmi,2019_Olmi,2019_Orlando,2020_Orlando} plasma physics~\citep{2022_Bodo,bugli2025}, planet formation~\citep{2021_Lesur,2021_Flock}, and solar physics~\citep{reville2015,perri2018}, to name just a few.

The new GPU-accelerated implementation of the code (\gpluto{}) is a complete C++ rewrite of the legacy code, which employs the OpenACC programming model to harness the vast computing power and energetic efficiency of modern GPU architectures, offering significant advantages in terms of computational speed, power consumption, and cost-effectiveness.
Switching to C++ is justified by the need to to take advantage of specific language extensions such as function templates, classes and polymorphism.
\gpluto{} keeps the strengths of the previous versions, such as stability, versatility, and user-friendliness, while also giving access to the most recent high-performance computing resources.
This work outlines the code's fundamental design principles and the remarkable results achieved in terms of simulation performance, while a more comprehensive description of the new code and all of its physical modules will appear in a future manuscript.


In this work we present the technical details of \gpluto{} by introducing its underlying algorithmic framework, data management and parallelization strategies, and implementation solutions. 
We also showcase a series of benchmark tests, highlighting the code's effectiveness in reproducing them as well as its notable performance.
This paper has a companion, complementary paper that focuses on the Lagrangian Particles module integration and its GPU offloading \citep{suriano2025}.
The paper is organized as follows: in \S\ref{sec:num_method}, we briefly recap the equations of MHD and the finite volume framework.
In \S\ref{sec:gPLUTO_design_for_GPUs}, we illustrate technical aspects concerning the OpenACC implementation, while in \S\ref{sec:num_tests}, we show the parallel performance on two standard use cases.
Finally, we draw our conclusions in \S\ref{sec:summary}.

\section{Numerical methods}
\label{sec:num_method}
%

\subsection{MHD equations}

Following its predecessor \citep{Mignone_PLUTO2007, Mignone_PLUTO2012}, \gpluto{} is designed to solve a variety of systems of conservation laws,
\begin{equation}\label{eq:cons_law}
\frac{\partial \cU}{\partial t} + \nabla \cdot \tens{T} = \cS \,,
\end{equation}
where $\cU$ is an array of conservative quantities, $\tens{T}$ is a flux tensor and $\cS$ defines the source terms.
Although different sets of equations may be solved, for the present purpose, we only focus on the single fluid classical MHD equations:
\begin{equation}\label{eq:mhd}
  \cU = \left( \begin{array}{c}
  \rho \\
  \rho \vec{v} \\
  \vec{B} \\
  \cE \\
\end{array}\right), \quad
 \tens{T} = \left( \begin{array}{c}
  \rho \vec{v} \\
  \rho \vec{v} \vec{v} - \vec{B B} + p_t \tens{I} \\
   \vec{v} \vec{B} - \vec{B} \vec{v} \\
   (\cE + p_t)\vec{v} - (\vec{v} \cdot \vec{B}) \vec{B} \\
 \end{array}\right)^T ,
\end{equation}
where $\rho$ is the fluid density, $\vec{v}$ is the fluid velocity, $\vec{B}$ is the magnetic field vector, $p_t = p + \vec{B}^2/2$ is the total (thermal + magnetic) pressure and
\begin{equation}\label{eq:Energy}
\cE = \frac{p}{\Gamma -1} + \frac{\rho \vec{v}^2}{2} + \frac{\vec{B}^2}{2} \,,
\end{equation}
is the total energy density comprising thermal, kinetic and magnetic contributions.
In Eq.~\ref{eq:Energy} we have assumed an ideal equation of state with adiabatic index $\Gamma$.

The divergence-free constraint, $\nabla \cdot \vec{B} = 0$, is not automatically preserved by the underlying discretization method and must be specifically enforced.
To this purpose, the code allows different algorithms to be selected, including include Powell's eight wave formulation \citep{Powell_etal1999}, divergence cleaning \citep{Dedner_etal2002, Mignone_etal2010} and the constrained transport method \citep[see][]{Evans_Hawley1988, Balsara_Spicer1999} in the more recent formulation by \citep{Mignone_DelZanna2021}.

Relativistic and non-ideal extensions of the previous equations will be described in a forthcoming work.

\subsection{Discretization}

We employ a Cartesian coordinate system with axes defined by the unit vectors $\hvec{e}_x = (1,0,0)$, $\hvec{e}_y =(0,1,0)$ and $\hvec{e}_z =(0,0,1)$. 
A uniform mesh with coordinate spacing $\Delta x$, $\Delta y$ and $\Delta z$ is assumed to cover the computational domain.

\gpluto{} employs a finite volume (FV) framework to evolve cell-averaged conservative variables in time, given specified initial and boundary conditions.
The FV method is based on the integral form of the governing differential equations, obtained by integrating Eq.~(\ref{eq:cons_law}) over a control volume corresponding to a single computational zone. 
Integrating over space within a single cell yields the semi-discrete form of the conservation equations:
\begin{equation}   \label{eq:FV}
  \begin{array}{lcl}
    \DS \frac{d\av{\cU}_{\cc}}{dt} = 
    &-& \DS \frac{\Delta t}{\Delta x}\left( 
      \hat{F}_{i+\HALF,j,k} - \hat{F}_{i-\HALF,j,k} \right) 
    \\ \noalign{\medskip}
    &-& \DS \frac{\Delta t}{\Delta y}\left( 
      \hat{G}_{i,j+\HALF,k} - \hat{G}_{i,j-\HALF,k} \right)    
    \\ \noalign{\medskip}
    &-& \DS \frac{\Delta t}{\Delta z}\left( 
      \hat{H}_{i,j,k+\HALF} - \hat{H}_{i,j,k-\HALF} \right)   \,,  
  \end{array}
\end{equation}
where $\cc = (i,j,k)$ is a short-hand notation for a single computational cell centered at $(x_i, y_j, z_k)$, $\{i,\,j,\,k\}$ is a triplet of zone indices in the three coordinate directions while cell interfaces are located at $\cc\pm\hvec{e}_x/2,\, \cc\pm\hvec{e}_y/2,\,$ and  $\cc\pm\hvec{e}_z/2$ in the three directions, respectively.
In Eq.~(\ref{eq:FV}), $\av{\cU}_\cc$ represents the volume-averaged conserved quantities within each cell, while $\hat{F}, \hat{G}$, and $\hat{H}$ are the surface-averaged numerical fluxes across the corresponding cell interfaces, computed via a Riemann solver.

At the numerical level, Eq.~(\ref{eq:FV}) is treated as a system of ordinary differential equations (ODEs) in time, and discretized using a strong-stability-preserving (SSP) Runge-Kutta (RK) method \citep{ShuOsher1988}.
For example, a third-order SSP Runge-Kutta scheme proceeds as:
\begin{equation}\label{eq:RK3}
  \begin{array}{lcl}
  \DS \av{\cU}^{*} &=& \DS \av{\cU}^{n} 
                                + {\cL}^n 
  \\ \noalign{\medskip}                                 
  \DS \av{\cU}^{**} &=& \DS 
      \frac{3\av{\cU}^n + \av{\cU}^{*} + \cL^{*}}{4}
  \\ \noalign{\medskip}                                 
  \DS \av{\cU}^{n+1} &=& \DS 
      \frac{\av{\cU}^n + 2\av{\cU}^{**} + \cL^{**}}{3} \,,
  \end{array}
\end{equation}
where, for brevity, the cell index subscript $\cc$ has been omitted. 
In the expression above, $\cL^n \equiv \cL(\av{\cU}^n)$ represents the right-hand side of Eq.~(\ref{eq:FV}), evaluated using the solution at time level $n$, and similarly for $\cL^{*}$ and $\cL^{**}$ at intermediate stages.

For each Runge-Kutta stage, the right-hand side $\cL$ is typically computed by first performing a reconstruction step, in which zone-centered primitive variables are extrapolated to cell interfaces. 
For instance, in the $x$-direction:

\begin{equation}\label{eq:reconstruct}
  \cV^L_{i+\HALF} = \cR^+_x(\cV_\cc)\,,\qquad
  \cV^R_{i-\HALF} = \cR^-_x(\cV_\cc)
\end{equation}
where $\cR^{\pm}_x$ denotes a non-oscillatory, high-order polynomial reconstruction operator (see, e.g., \citep{Colella_Woodward1984, Suresh_Huynh1997, Borges_WENOZ2008}), applied in the positive ($+$) or negative ($-$) $x$-direction.

Here, $\cV_\cc = \{\rho,\, \vec{v},\,\vec{B},\, p\}$ represents the array of primitive variables (density, velocity, magnetic field, and pressure), which are preferred over conservative variables during the reconstruction process.
This choice is made because primitive variables typically exhibit smoother variations across the computational domain — except at shocks or contact discontinuities — thereby reducing numerical errors and mitigating spurious oscillations near sharp gradients.
In MHD, primitive variables are recovered from the conservative ones via straightforward inversion:
\begin{equation}
    \vec{v} = \frac{\rho\vec{v}}{\rho}\,,\qquad
    p = (\Gamma-1)\left(\cE - \frac{\rho\vec{v}^2}{2}
                            - \frac{\vec{B}^2}{2}\right)
\end{equation}
where $\Gamma$ is the adiabatic index, and $\cE$ is the total energy density.
Employing primitive variables during reconstruction helps limit the occurrence of unphysical states, such as negative pressures, especially in regions near strong gradients.

The reconstructed values at cell interfaces serve as input states to the Riemann solver, which computes a stable, upwinded numerical flux. 
For example:
\begin{equation}
     \hat{F}_{i+\HALF} 
   =   \frac{F(\cV^L_{i+\HALF}) + F(\cV^R_{i+\HALF})}{2} 
     - {\cal D}_{i+\HALF}\cdot \left(\cU^R_{i+\HALF}- \cU^L_{i+\HALF}\right)
\end{equation}
where ${\cal D}$ is a dissipation matrix whose form depends on the specific Riemann solver employed. 
The subscripts $j$ and $k$ have been omitted here for clarity, as the context is now unambiguous.

Magnetic fields can be evolved either in a zone-centered formalism, as described so far, or using the constrained transport (CT) method \citep{Evans_Hawley1988, Londrillo_DelZanna2004}. 
The advantage of CT is its ability to maintain the divergence-free condition for the magnetic field to machine precision throughout the simulation.
Further details on the numerical implementation can be found in the original \pluto{} code paper \citep{Mignone_PLUTO2007}, and in recent developments related to CT techniques in \citep{Mignone_DelZanna2021}.

As its predecessor, \gpluto{} offers a variety of algorithms with different levels of accuracy and computational cost to choose from. In the present work, we employ the third-order SSP Runge-Kutta time-stepping scheme (Eq.~(\ref{eq:RK3})), combined with either parabolic reconstruction \citep{Mignone2005} or the WENOZ reconstruction method \citep{Borges_WENOZ2008}.

\section{PLUTO design for GPUs}
\label{sec:gPLUTO_design_for_GPUs}
%
%
Modern Graphics Processing Units (GPUs) are specialized, highly parallel processors that have become cornerstones of high-performance scientific computing. Their computational power originates from an architecture containing hundreds or thousands of simple cores designed to execute the same instruction simultaneously across large sets of data. This paradigm makes GPUs exceptionally efficient for algorithms characterized by high data parallelism, such as the mesh-based solvers used in our work. Computations are offloaded from the host CPU to the GPU (called in this context device or accelerator) by launching functions called kernels, which are managed and executed by the device's hardware scheduler. The execution is organized hierarchically. A thread is a single stream of instructions of that runs in parallel with other threads and is mapped to the GPU cores.  Individual threads that execute the same instruction are grouped into warps and multiple warps are contained in blocks, that are scheduled for execution on the GPU's fundamental hardware units, the Streaming Multiprocessors (SMs). Each SM is an independent multiprocessor containing its own execution cores, scheduler, and a high-speed shared memory, allowing it to execute one or more thread blocks concurrently.

\subsection{OpenACC}

The programming model chosen for GPU porting of PLUTO is OpenACC, a simpler and higher-level alternative to CUDA. OpenACC is a directive-based programming model for GPU computing developed by Cray, CAPS, NVIDIA, and PGI in 2011.
It relies on information provided by the programmer to the compiler through directives, library routines, and environment variables. 
These annotations instruct the compiler in porting the code to the device.
The complex and low-level aspect of the parallelization is left to be handled by the compiler, facilitating the user’s work. 
Another advantage of choosing a directive-based approach is interchangeability, an important feature allowing the code to be compiled and run, ignoring the OpenACC directives, making it suitable for both CPU and GPU computing.
For every GPU-accelerated application, the computing work is divided among CPUs and GPUs. 
In this context, \quotes{host} refers to the primary processor in a system (the CPU) while the \quotes{device} refers to the secondary processor (the GPU) responsible for the acceleration.

In broad terms, in \gpluto{} the entire computational workload is handled by the GPUs, while the CPUs oversee overall system operations, including running the operating system, managing system memory, and handling I/O operations.
The reason for this strict division is performance. 
Executing all the computations on the device permits, in fact, the exploitation of its potential without introducing data movement between the memories of the two processors. 
The alternate use of host and device would introduce memory copies to assure coherence between data and, hence, correct results. OpenACC directives are indicated with the keyword \Cinline{pragma} followed by the \Cinline{acc} sentinel, the directive itself, and optional clauses:
\begin{Ccode}
#pragma acc < directive > < clause , clause >
\end{Ccode}
The directive is applied to the instruction, loop, or structured code block that follows.
The base approach when accelerating an application with OpenACC is using compute directives (we mainly use \Cinline{parallel loop}) that indicate to the compiler regions of the code to execute on the device, the kernels. 
Data directives, on the other hand, such as \Cinline{enter data copyin} and \Cinline{exit data delete}, permit explicit control of the data movement. 
This is crucial since, in many cases, the performance of the kernels is limited by the memory bandwidth of the device, and thus, reducing the amount of data needed to be moved gives a huge benefit.

\subsection{Compute Directives} 
\label{sec:compute_directive}
A typical compute construct of \gpluto{} is reported in Fig.~\ref{fig:compute}, where \Cinline{NX}, \Cinline{NY}, \Cinline{NZ} define domain size, \Cinline{data->Vc} is the array of the primitive quantities (i.e., density, pressure and velocities), that contains \Cinline{NVAR} physical quantities for each point of the 3-dimensional domain.
\begin{figure}[htbp]
\centering
\begin{ccode}
#pragma acc parallel loop collapse(3) present(data)
for (k = 0; k < NZ; k++){
for (j = 0; j < NY; j++){
for (i = 0; i < NX; i++){
  ...
  #pragma acc loop seq
  for (nv = 0; nv < NVAR; nv++){
    data->Vc(i,j,k,nv) = ...;
  } 
  ...
}}}
\end{ccode}
\caption{Kernel accelerated with a compute directive.}
\label{fig:compute}
\end{figure}

The first directive is at the core of the acceleration of the application: \Cinline{#pragma acc parallel loop} directive tells the compiler to parallelize the loop, distributing its iterations across the threads of the target accelerator. 
The loop body is then executed concurrently, significantly improving the performance of the code.
The \Cinline{collapse(3)} clause applies in this case to three tightly nested loops. 
Hence, the compiler treats the construct as a single loop of \Cinline{NX x NY x NZ} iterations. 
This makes it easier for the compiler to map the construct onto the device's computing resources, potentially improving parallelization efficiency.
The \Cinline{present} clause implies that data is already present on the GPU and that a copy from the main memory is not necessary. 
If this is not true a fatal error at the execution would be generated.

The innermost loop, over \Cinline{NVAR} elements, is marked with \Cinline{seq}, meaning that the operation on different values of \Cinline{nv} is executed sequentially by the same thread. 
Note that while parallelization of this inner loop could be possible, it usually tends to lower the performance, as the number of variables \Cinline{NVAR} is typically much smaller than the number of grid points.
Moreover, having the threads corresponding to loops with many iterations makes sense for optimal utilization and memory access (see Sec.~\ref{sec:memory}).

\subsection{Memory Management} \label{sec:memory}

To manage data transfer to the device in \gpluto{}, we combine explicit OpenACC data directives with CUDA Unified Memory, enabled via the compile-time flag \Cinline{-gpu=managed}. Unified Memory provides a shared address space accessible by both the CPU and GPU through a single pointer, making it particularly suitable for dynamically allocated data. In most cases, the NVIDIA driver automatically handles data movement. However, we employ explicit data directives to ensure correct deep copying of structured and nested data that are widely used within the code.

GPUs employ various techniques to hide memory latency that can represent an important performance bottleneck. 
Most of these techniques, such as warp scheduling\footnote{When some threads within a warp need to access memory, the scheduler can switch to other warps with active threads, allowing the GPU to continue executing instructions without waiting for the memory access to complete.} and prefetching\footnote{GPUs can predict and preload data into caches before it is actually needed, reducing the time it takes to fetch the data when it is required by the computation.} are handled by the device itself but the programmer can actively act on other important aspects such as coalesced memory access.

\paragraph{Coalesced memory access} Coalesced memory access is a memory access pattern in parallel computing, particularly on GPUs, where multiple threads access consecutive or contiguous memory locations in a single memory transaction. 
This pattern is essential for achieving efficient memory bandwidth utilization and improving the overall performance of parallel applications. 
In the pseudo-code of Fig.~\ref{fig:compute}, coalesced memory access is achieved, granting optimal parallelization. 
Here, consecutive values of the index \Cinline{i} correspond to consecutive threads but also to adjacent memory access (\Cinline{i} is the fastest index of \Cinline{data->Vc}). 
The advantage of this memory pattern can be traced back to the thread memory access. 
When a thread needs data for an operation, it receives access to many adjacent memory locations. 
Neighbor threads can be exploited immediately if they need data received in this memory transaction, hence the advantage of them accessing adjacent memory locations.

\paragraph{Ary classes} Optimal memory access pattern is one of the main reasons behind the profound change of many arrays of the code together with the migration to \texttt{C++}.  
We implemented classes called \Cinline{Ary} to allocate and handle memory access for multi-dimensional arrays. In the details, we moved from a pointer to pointer approach to one-pointer, flat arrays using classes to hide index computations.
Additionally, element ordering can be changed whether or not we're compiling with OpenACC or based on the array type, covering the complexity of this operation in the class itself.
We now discuss a technique we adopted widely, using the \Cinline{Ary} class with flat arrays to improve memory access. 

In Fig.~\ref{fig:2d} \Cinline{rhs} is a 2D array of size \Cinline{NX x NVAR} created from the single, contiguous block of memory \Cinline{d->sweep.rhs} of size \Cinline{NX x NY x NZ x NVAR}, based on an offset defined by \textit{j} and \textit{k}. 
This assures that each thread writes on a unique memory address (privatization\footnote{Thread privatization is the process of giving each thread within a kernel its own private copy of a variable, ensuring that parallel computations don't interfere with each other by creating race conditions.}), which is necessary to grant correct results. 
Moreover, class definition assures that neighboring threads corresponding to consecutive values of \textit{i} access adjacent memory locations. 

\begin{figure}[htbp]
    \centering
    \begin{ccode}
#pragma acc parallel loop collapse(2)
for (k = 0; k < NZ; k++){
for (j = 0; j < NY; j++){
  long int offset = NVAR*NX*(j + NY*k);
  Ary2D flux(&d->sweep.flux[offset],NX,NVAR);
  Ary2D rhs(&d->sweep.rhs[offset],NX,NVAR);

  #pragma acc loop vector
  for (i = 0; i < NX; i++){
    #pragma acc loop seq
    NVAR_LOOP(nv) {
      rhs(i,nv) = -dtdx*(flux(i+1,nv) - flux(i,nv)); 
    } 
}}}
\end{ccode}
\caption{Computation of the right-hand side contribution along the x-direction.}
\label{fig:2d}
\end{figure}

Then, when computing the y-direction contribution, the same technique is applied to the same memory pool \Cinline{d->sweep.rhs}, changing the order of write and access to the elements to obtain again optimal access pattern again.
Other arrays such as \Cinline{Vc} and \Cinline{Uc} have fixed ordering: \Cinline{i,j,k,nv} in order of decreasing index speed.

\paragraph{Template function} Another useful feature of \texttt{C++} that has been adopted is function templates. Many kernels operate on arrays whose indexing depends on the direction of integration, which can vary at runtime. This runtime variability prevents the compiler from knowing in advance which registers will be needed, often resulting in these variables being placed in slower memory spaces such as local or shared memory instead of registers. By using templated functions, we can specialize the code based on template parameters—in our case, the integration direction \texttt{dir}. This enables the compiler to inline the specialized functions and propagate constant values, eliminating branches and improving performance. Fig.~\ref{fig:hlld_2} shows an example of the Riemann solver with \texttt{dir} as a template parameter. In this case, the array \texttt{v} can be allocated in registers at compile time.

\begin{figure}[htbp]
\centering
\begin{ccode}    
template<int dir>
void Riemann_Solver (Data *d, Grid *grid, RBox *box);

void Riemann_Solver_T (Data *d, Grid *grid, RBox *box){
  if(dir==IDIR) Riemann_Solver<IDIR>(d, grid, box);
  else if(dir==JDIR) Riemann_Solver<JDIR>(d, grid, box);
  else if(dir==KDIR) Riemann_Solver<KDIR>(d, grid, box);
}

template<int dir>
void Riemann_Solver (Data *d, Grid *grid, RBox *box)
{
  int VXn = VX1+dir;

   #pragma acc parallel loop collapse(2)
   for (k = kbeg; k <= kend; k++){
   for (j = jbeg; j <= jend; j++){
     #pragma acc loop
     for (i = ibeg; i <= iend; i++){
       int v[NVAR];
       v[VXn] = ...;
     }
    }}
}
\end{ccode}
    \caption{\small Templated Riemann Solver. Three version of the routine are compiled with different values of the direction of integration \texttt{dir}: \texttt{IDIR}, \texttt{JDIR}, \texttt{KDIR}}
    \label{fig:hlld_2}
\end{figure}

\subsection{Multi GPU communication}

Multi-process communication at each time step is necessary in \gpluto{} to assure correct results when the domain is split between processes. 
Communication is based on the message passing interface (MPI) protocols. 
While in \pluto{} the data exchange was based on the library ArrayLib (\citep{mignoneaoptimization}), it has here been rewritten using a more basic scheme based on standard buffer arrays and \verb|MPI_Isend|/\verb|MPI_Irecv| calls. 
To exploit the characteristics of GPUs, we employ an asynchronous approach for data exchange. 
This requires each processor to communicate potentially simultaneously with its 8 neighbors in 2D or 26 neighbors in 3D (this includes also nearby processors in the diagonal directions).
The pseudo-code in Fig.~\ref{fig:par_ex} describes the dedicated routine, considering one direction of exchange (\Cinline{IDIR}).

\begin{figure}[htbp]
    \centering
    \begin{ccode}
MPI_Irecv (recv_buf_2, ..., nrnks[IDIR][1], ...);
MPI_Irecv (recv_buf_1, ..., nrnks[IDIR][0], ...);

#pragma acc parallel loop collapse(4) async(1)
LOOP(...){
  send_buf_1[ind] = Vc((ibeg+i),j,k,nv);
}

#pragma acc parallel loop collapse(4) async(2)
LOOP(...){
  send_buf_2[ind] = Vc((iend-(n1g-1)+i),j,k,nv);
}

#pragma acc wait

MPI_Isend(send_buf_1, ..., nrnks[IDIR][0], ...);
MPI_Isend(send_buf_2, ..., nrnks[IDIR][1], ...);

MPI_Waitall(..., &recv[2], ...);
MPI_Waitall(..., &recv[1], ...);

#pragma acc parallel loop collapse(4) async(1)
LOOP(...){
  Vc(i,j,k,nv) = recv_buf_1[ind];
}

#pragma acc parallel loop collapse(4) async(2)
LOOP(...){
  Vc((iend+1+i),j,k,nv) = recv_buf_2[ind];
}

#pragma acc wait

MPI_Waitall(..., &send[1], ...);
MPI_Waitall(..., &send[2], ...);
\end{ccode}
\caption{Asynchronous implementation of the parallel data exchange routine.}
\label{fig:par_ex}
\end{figure}

Initially, the code initializes non-blocking receives (\verb|MPI_Irecv|) to get data from neighboring processes. 
Then, data to be sent to neighboring processes are packed into buffers allocated with \verb|acc_malloc|\footnote{OpenACC function that allocates memory directly in the accelerator's global memory.}, and each kernel has an \verb|async| clause with a different number: each kernel is independent, and they could ideally run at the same time on distinct streams (sequences of operations). 
After the code initializes non-blocking send calls (\verb|MPI_Isend|) to send data to neighboring processes and then it waits for all receive operations to complete. 
When this happens, the received data is unpacked into the appropriate variables, again concurrently for each \verb|recv_buf|. 
Finally, it waits for all non-blocking send operations to complete and ensures all asynchronous operations on the GPUs are terminated before proceeding.

\section{Numerical tests and Discussion}
\label{sec:num_tests}
%
%

\begin{table*}[!bp] 
\centering
\begin{tabularx}{\textwidth}{@{}lXXXX@{}}
\toprule
 & \multicolumn{2}{c}{\textbf{Leonardo}} & \textbf{MeluXina} & \textbf{MareNostrum 5} \\
\cmidrule(lr){2-3} 
\textbf{Detail} & \textbf{Booster} & \textbf{DCGP} & \textbf{GPU} & \textbf{ACC} \\
\midrule
Proprietary Company & \multicolumn{2}{c}{CINECA} & LuxProvide & BSC \\
\addlinespace
Active from year & \multicolumn{2}{c}{2022} & 2021 & 2023 \\
\addlinespace
Node Model & BullSequana X2135 & BullSequana X2140 & BullSequana XH2000 & BullSequana XH3000 \\
\addlinespace
CPUs (per node) & Intel Ice Lake 8358 & 2x 56-core Intel Sapphire Rapids 8480+ & 2x AMD EPYC Rome 7452 & 2x Intel Sapphire Rapids 8460Y+ \\
\addlinespace
GPUs (per node) & 4x NVIDIA Ampere (64 GB HBM2e) & N/A (CPU-only) & 4x NVIDIA Ampere (40 GB HBM) & 4x NVIDIA Hopper (64 GB HBM2) \\
\bottomrule
\end{tabularx}%
\caption{Summary of computational hardware resources utilized.}
\label{tab:hardware_specs}
\end{table*}

Here, we present numerical benchmarks conducted on CPU and GPU based partitions of three different HPC platforms: MareNostrum 5 Accelerated Partition (GPU), Leon\-ardo Booster (GPU), MeluXina GPU nodes (GPU), and Leonardo DCGP (CPU).
Every run is performed employing the entire computational power of each node, meaning 4 GPUs per node on GPU partitions and 112 CPU-cores on Leonardo DCGP.

Leonardo is hosted by Cineca and has two main partitions: Booster module and Data-centric General Purpose (DCGP) module. 
The Booster partition is equipped with Atos BullSequana X2135 \quotes{Da Vinci} single-node GPU blade with four NVIDIA Ampere GPUs/node (64\,GB HBM2e). 
The DCGP module has instead Atos BullSequana X2140 three-node CPU blade, with dual socket 56-core Intel Sapphire Rapids Intel Xeon Platinum 8480+.
MeluXina belongs to LuxProvide and the GPU nodes provide Atos BullSequana XH2000 platform with four NVIDIA Ampere GPUs (40\,GB HBM).
Finally, MareNostrum 5 is at the Barcelona Supercomputing Center (BSC), and the  Accelerated Partition (ACC) we used is a BullSequana XH3000 model with four NVIDIA Hopper GPUs (64\,GB HBM2). See Tab.~\ref{tab:hardware_specs} for a summary of the systems specifics.

\subsection{3D Orszag Tang Vortex}
\label{subsec:orszag_tang_test}

The first benchmark is a 3D version of the well-known Orszag-Tang test problem, with a similar to that of \citep{Helzel_etal2011}.
We consider a periodic Cartesian domain of size $L_x = L_y = L_z = 1$ with constant initial pressure $p = 5/3$ and density $\rho = 25/9$. 
Velocity and magnetic field are initialized as follows
\begin{equation}
  \begin{array}{l}
  \vec{v} = - \zeta(z)\sin(2\pi y) \hvec{e}_x 
                  + \zeta(z)\sin(2\pi x) \hvec{e}_y 
                  + 0.2 \sin(2\pi z)\hvec{e}_z \;,  
  \\ \noalign{\medskip}
  \vec{B} = - \sin(2\pi y)\hvec{e}_x 
                       + \sin(4\pi x)\hvec{e}_y \;,
  \end{array}
\end{equation}
where $\zeta(z) = 1 + \sin(2\pi z)/5$.
Computations are stopped at $t = 0.5$ and the test employs the HLLD Riemann solver \citep{Miyoshi_Kusano2005}, WENOZ reconstruction \citep{Borges_WENOZ2008}, and the RK3 time-stepping scheme.
All computations are performed in double-precision arithmetic.
\paragraph{Weak scaling} 

\begin{table}[H]
\centering
\begin{tabularx}{\columnwidth}{l X}
\toprule
\textbf{System} & \textbf{Resolution per Node} \\
\midrule
Leonardo Booster & 704 x 704 x 352 \\
Leonardo DCGP & 704 x 704 x 350 \\
MareNostrum 5 ACC         & 704 x 704 x 352 \\
MeluXina GPU nodes        & 512 x 512 x 512 \\
\bottomrule
\end{tabularx}
\caption{Weak scaling resolution (per node) for the 3D Orszag-Tang test.}
\label{tab:weak_nov}
\end{table}

We first assess the parallel performance using weak scale metric, implying that grid resolution per node - reported in Tab. \ref{tab:weak_nov} - is kept constant.
For these runs, we employ the constrained transport method (with UCT HLLD average, \citep{Mignone_DelZanna2021}) to enforce the divergence-free condition of magnetic field.
On the GPU partitions, we chose a resolution that maximized device occupancy to achieve the best performance. A $704 \times 704 \times 352$ resolution was used to nearly fill the device memory on both Leonardo Booster and Marenostrum 5 ACC. A lower resolution was necessary for MeluXina to accommodate its smaller GPU memory. The runs on Leonardo DCGP used a slightly different resolution to allow for domain decomposition across 112 processes while maintaining a nearly identical computational load per process as: $\frac{N_{\rm t, CPU node}}{N_{\rm t, GPU node}}\sim 0.99$, with $N_{\rm t}$ the total number of grid points.
%

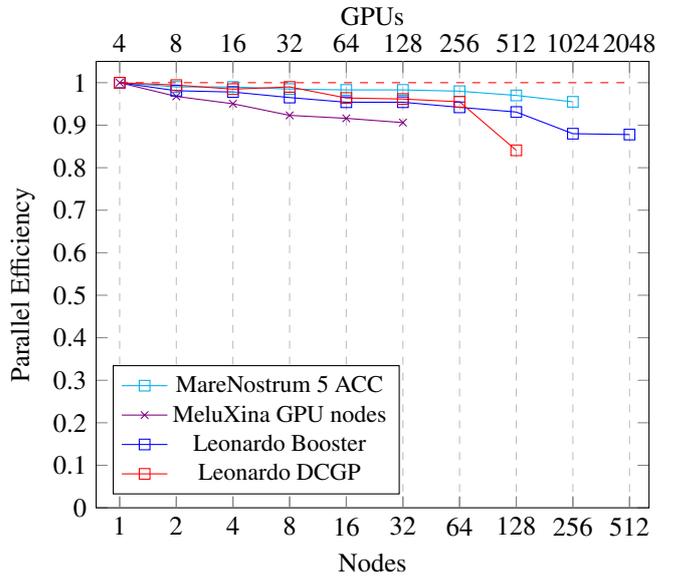
\begin{figure}[htbp]
\centering
\begin{tikzpicture}
\begin{axis}[
    width=\columnwidth,  
    height=7.5cm,  
    legend style={nodes={scale=0.9, transform shape}},
    xlabel={Nodes},
    ylabel={Parallel Efficiency},
    xmin=0.75, xmax=650,
    ymin=0., ymax=1.05,
    xtick={1,2,4,8,16,32,64,128,256,512},
    xticklabels={1, 2, 4, 8, 16, 32, 64, 128, 256, 512},
    ytick={0.0,0.1,0.2,0.3,0.4,0.5,0.6,0.7,0.8,0.9,1.0},
    legend pos=south west,
    grid style=dashed,
    xmode=log,
    log basis x=2,
    extra x ticks={1,2,4,8,16,32,64,128,256,512},
    extra x tick labels={4,8,16,32,64,128,256,512,1024,2048}, 
    extra x tick style={grid=major, xticklabel pos=upper}, 
]
\addplot[color=cyan, mark=square]
    coordinates {
    (1,1.0)(2,0.99)(4,0.99)(8,0.985)(16,0.983)(32,0.983)(64,0.98)(128,0.97)(256,0.955)};
    \addlegendentry{MareNostrum 5 ACC}
\addplot[color=violet, mark=x]
    coordinates {
    (1,1.0)(2,0.9677)(4,0.9504)(8,0.9231)(16,0.91613)(32,0.90606)};
    \addlegendentry{MeluXina GPU nodes}
\addplot[color=blue, mark=square]
    coordinates {
    (1,1.0)(2,0.981)(4,0.978)(8,0.965)(16,0.954)(32,0.954)(64,0.942)(128,0.931)(256,0.88)(512,0.878)};
    \addlegendentry{Leonardo Booster}
\addplot[color=red, mark=square]
    coordinates {
    (1,1)(2,0.9943475564)(4,0.9847872268)(8,0.9900110223)(16,0.9639790716)(32,0.9615925059)(64,0.9550106326)(128,0.8409117407)};
    \addlegendentry{Leonardo DCGP}
\addplot[color=red, dashed]
    coordinates {
    (1,1.0)(2,1.0)(4,1.0)(8,1.0)(16,1.0)(32,1.0)(64,1.0)(128,1.0)(256,1.0)(512,1.0)
    };
\end{axis}
\node[anchor=south, yshift=10pt] at (current axis.north) {GPUs};
\end{tikzpicture}
    \caption{Weak scaling results for the 3D Orszag-Tang on MareNostrum 5 ACC, MeluXina GPU nodes and Leonardo (Booster and DCGP).}
    \label{fig:weak_nov}
\end{figure}
Fig.~\ref{fig:weak_nov} presents the parallel efficiency expressed as the ratio of the wall clock times on $1$ node and $N$ nodes, $T_{1}/T_{N}$.
We have repeated the test on all 4 HPC systems up to 512 nodes (when permitted).
Among the tested systems, MareNostrum 5 ACC delivered the best results, maintaining a parallel efficiency above $95\,\%$ even at 256 nodes. 
On Leonardo Booster, we scaled up to 512 nodes, achieving $88\,\%$ efficiency at the largest scale. 
In contrast, MeluXina GPU nodes showed a steeper efficiency drop, reaching $91\,\%$ at just 32 nodes. 
The CPU runs on Leonardo DCGP maintain an efficiency above $95\,\%$ for up to 64 nodes, but this value drops to $84\,\%$ at 128 nodes. Further investigation of the reasons for this low efficiency at 128 nodes will be needed.

\begin{figure}[htbp]
\centering
\begin{tikzpicture}
\begin{axis}[
    width=\columnwidth,  
    height=7.5cm,  
    legend style={nodes={scale=0.9, transform shape}},
    xlabel={Nodes},
    ylabel={Performance (cell/sec/nodes)},
    ymin=6^6, ymax=10.5^9,
    ytick={10^6,10^7,10^8,10^9},
    xmin=0.75, xmax=650,
    xtick={1,2,4,8,16,32,64,128,256,512},
    xticklabels={1, 2, 4, 8, 16, 32, 64, 128, 256, 512},
    legend pos=south west,
    grid style=dashed,
    ymode=log,
    xmode=log,
    log basis x=2,
    extra x ticks={1,2,4,8,16,32,64,128,256,512},
    extra x tick labels={4,8,16,32,64,128,256,512,1024,2048}, 
    extra x tick style={grid=major, xticklabel pos=upper}, 
]
\addplot[color=cyan, mark=square]
    coordinates {
    (1,151126654.7)(2,150083303.5)(4,149980082.5)(8,148898418.5)(16,148568730.7)(32,148581384)(64,148111499.1)(128,146203085.7)(256,141806000.4)};
    \addlegendentry{MareNostrum 5 ACC}
\addplot[color=violet, mark=x]
    coordinates {
    (1,116583541.3)(2,112820453.2)(4,110805978.1)(8,107626985.1)(16,106806584.5)(32,105632572.6)};
    \addlegendentry{MeluXina GPU nodes}
\addplot[color=blue, mark=square]
    coordinates {
    (1,111831302.6)(2,109721278)(4,109377324.1)(8,108022806.2)(16,106701426.3)(32,106701426.3)(64,105411983.1)(128,104153332.5)(256,98563181.92)(512,98285539.15)};
    \addlegendentry{Leonardo Booster}
\addplot[color=red, mark=square]
    coordinates {
    (1,12025344.89)(2,11957372.3)(4,11886904.68)(8,11949958.67)(16,11635739.2)(32,11606932.08)(64,11527485.38)(128,10112253.7)};
    \addlegendentry{Leonardo DCGP}
\end{axis}
\node[anchor=south, yshift=10pt] at (current axis.north) {GPUs};
\end{tikzpicture}
    \caption{Performance of Orszag Tang weak scaling tests on MareNostrum 5 ACC, MeluXina GPU nodes and Leonardo (Booster and DCGP).}
    \label{fig:perf_nov}
\end{figure}
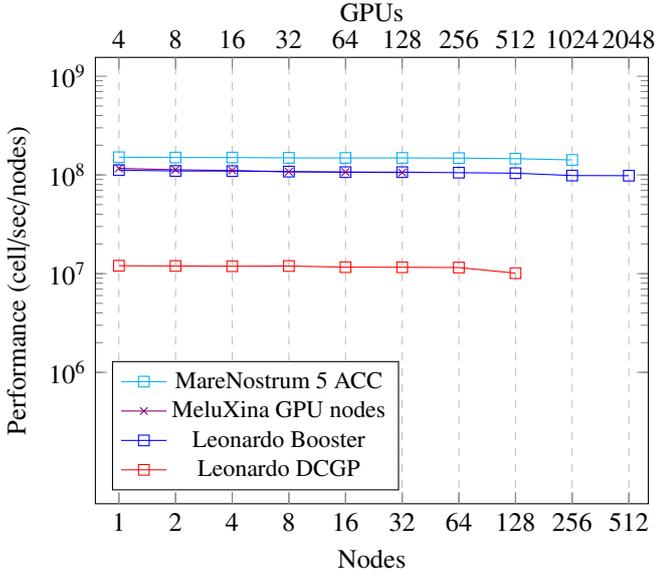

%
%
\begin{table}[htbp]
\centering
\begin{tabular}{ p{1.1cm}p{1.8cm}p{1.8cm}p{1.95cm} }
 \toprule
 Nodes & $T_{\text{GPUs}}$ (sec) & $T_{\text{CPUs}}$ (sec) & Acceleration \footnotesize $(T_{\text{CPUs}}/T_{\text{GPUs}})$ \\
 \midrule
1   & 312 & 2885 & 9.24 \\
2   & 318 & 2901.4 & 9.12 \\
4   & 319 & 2918.6 & 9.14 \\      
8   & 323 & 2903.2 & 8.98 \\
16  & 327 & 2981.6 & 9.11 \\
32  & 327 & 2989 & 9.14 \\             
64  & 331 & 3009.6 & 9.02 \\
128 & 335 & 3430.8 & 10.24 \\
 \bottomrule
\end{tabular}
\caption{CPU-GPU comparison of Orszag Tang weak scaling tests (Fig.~\ref{fig:weak_nov}) on Leonardo DCGP (4 A100 GPUs per node) and Booster partition (112 Intel Xeon 8480+ CPUs cores per node), respectively.}
\label{tab:comparison}
\end{table}
In Fig.~\ref{fig:perf_nov}, we show the corresponding performance of the weak scaling runs, measured in terms of the number of cells updated per second per node (computed as the number of grid points multiplied by the number of time steps divided by the execution time and the number of nodes). 
Note that CPU runs on Leonardo DCGP were approximately 9 times slower than GPU runs on the Leonardo Booster partition. 
Tab.~\ref{tab:comparison} provides the acceleration factors for the two types of runs.

\begin{figure}[htbp]
\centering
\begin{tikzpicture}
\begin{axis}[
    width=\columnwidth,  
    height=7.5cm,  
    legend style={nodes={scale=0.9, transform shape}},
    legend style={fill=white, fill opacity=0.8},
    xlabel={Nodes},
    ylabel={Speedup},
    xmin=0, xmax=340,
    ymin=0.85, ymax=9,
    xtick={1,2,4,8,16,32,64,128,256},
    xticklabels={1, 2, 4, 8, 16, 32, 64, 128, 256},
    ytick={1,2,4,8},
    yticklabels={1, 2, 4, 8},
    legend pos=north west,
    legend style={at={(0.03,0.86)}}, 
    grid style=dashed,
    xmode=log,
    log basis x=2,
    ymode=log,
    log basis y=2,
    extra x ticks={1,2,4,8,16,32,64,128,256,512},
    extra x tick labels={4,8,16,32,64,128,256,512,1024,2048},
    extra x tick style={grid=major, xticklabel pos=upper},
]

\addplot[color=cyan, mark=square] coordinates {(1,1.0)(2,1.989)(4,3.87)(8,7.56)};
\addlegendentry{MareNostrum 5 ACC}

\addplot[color=blue, mark=square] coordinates {(1,1.0)(2,1.96)(4,3.87)(8,7.46)};
\addlegendentry{Leonardo Booster}

\addplot[color=cyan, mark=square] coordinates {(2,1.0)(4,1.954)(8,3.85)(16,7.57)};
\addplot[color=cyan, mark=square] coordinates {(4,1.0)(8,1.99)(16,3.30)(32,7.49)};
\addplot[color=cyan, mark=square] coordinates {(8,1.0)(16,1.99)(32,3.87)(64,7.48)};
\addplot[color=cyan, mark=square] coordinates {(16,1.0)(32,1.85)(64,3.63)(128,7.08)};
\addplot[color=cyan, mark=square] coordinates {(32,1.0)(64,1.89)(128,3.89)(256,6.2)};

\addplot[color=blue, mark=square] coordinates {(2,1.0)(4,1.98)(8,3.85)(16,7.32)};
\addplot[color=blue, mark=square] coordinates {(4,1.0)(8,1.97)(16,3.76)(32,6.9)};
\addplot[color=blue, mark=square] coordinates {(8,1.0)(16,1.94)(32,3.76)(64,6.96)};
\addplot[color=blue, mark=square] coordinates {(16,1.0)(32,1.91)(64,3.78)(128,6.97)};
\addplot[color=blue, mark=square] coordinates {(32,1.0)(64,1.93)(128,3.76)(256,7.05)};

\addplot[color=red, dashed] coordinates {(0,0)(1,1.0)(2,2)(4,4)(8,8)};
\addplot[color=red, dashed] coordinates {(2,1)(4,2)(8,4)(16,8)};
\addplot[color=red, dashed] coordinates {(4,1)(8,2)(16,4)(32,8)};
\addplot[color=red, dashed] coordinates {(8,1)(16,2)(32,4)(64,8)};
\addplot[color=red, dashed] coordinates {(16,1)(32,2)(64,4)(128,8)};
\addplot[color=red, dashed] coordinates {(32,1)(64,2)(128,4)(256,8)};
\end{axis}
\node[anchor=south, yshift=10pt] at (current axis.north) {GPUs};
\end{tikzpicture}
\caption{Strong scaling tests of Orszag Tang on MareNostrum 5 ACC and Leonardo Booster.}
\label{fig:strong_nov}
\end{figure}
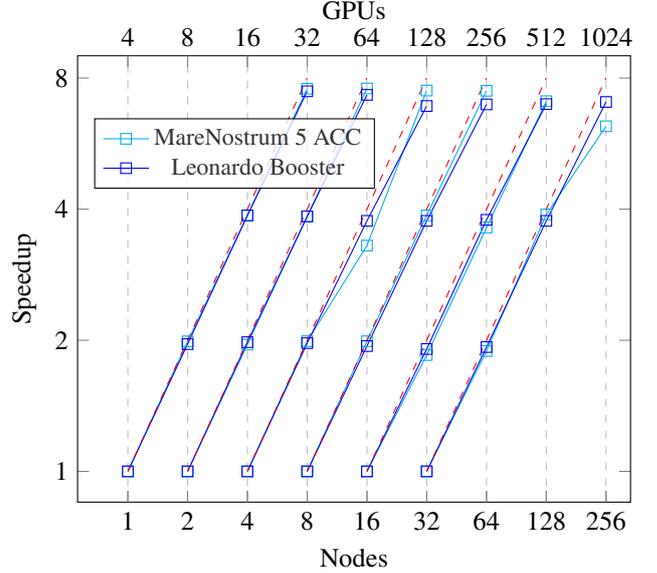

%
\paragraph{Strong scaling} We then present, in Fig.~\ref{fig:strong_nov}, strong scaling performance results of \gpluto{}, obtained by repeating the test using the divergence cleaning method for the control of the magnetic field divergence. 
This allowed for choosing a slightly larger grid than that used in the weak scaling test because this algorithm requires less device memory than the constrained transport method (Tab.~\ref{tab:scaling_efficiency_combined}).
In this test, the domain size remains fixed while the number of processes varies. 
As a result, GPU utilization is optimal in the initial runs but degrades, along with performance, as more nodes are employed. 
To address this effect, we performed six sets of four runs each, maintaining a fixed resolution as detailed in Tab.~\ref{tab:scaling_efficiency_combined}. 
The resolution was selected to ensure that the grid size nearly filled the GPU VRAM at the lowest node count within each group. 
We then successively doubled the number of nodes and processes four times per group, obtaining strong scaling efficiencies.
The results obtained on MareNostrum 5 ACC and Leonardo Booster largely overlap, with average efficiencies of the runs at higher numbers of nodes of each group at $0.90$ and $0.89$, respectively. 
Tab.~\ref{tab:scaling_efficiency_combined} provides the exact scaling efficiencies, computed by comparing the measured values of speedup, plotted in Fig.~\ref{fig:strong_nov}, to the ideal ones that correspond to the red lines in figure (speedup=$1$,$2$,$4$,$8$).
\begin{table}[htbp]
\centering
\begin{tabular}{ccc}
\toprule
\textbf{Nodes} & \textbf{MareNostrum Efficiency} & \textbf{Leonardo Efficiency} \\ \midrule
\multicolumn{3}{c}{\textbf{Group 1, Base: 1 Node, Resolution: 832 x 832 x 416}} \\ \midrule
1              & 1.00                & 1.00                    \\
2              & 0.98                & 0.98                    \\
4              & 0.97                & 0.97                    \\
8              & 0.94                & 0.93                    \\ \midrule
\multicolumn{3}{c}{\textbf{Group 2, Base: 2 Nodes, Resolution: 832 x 832 x 832}} \\ \midrule
2              & 1.00                & 1.00                    \\ 
4              & 0.99                & 0.99                    \\ 
8              & 0.96                & 0.96                    \\ 
16             & 0.94                & 0.92                    \\  \midrule
\multicolumn{3}{c}{\textbf{Group 3, Base: 4 Nodes, Resolution: 1664 x 832 x 832}} \\ \midrule
4              & 1.00                & 1.00                    \\ 
8              & 0.99                & 0.99                    \\ 
16             & 0.83                & 0.94                    \\ 
32             & 0.94                & 0.86                    \\  \midrule
\multicolumn{3}{c}{\textbf{Group 4, Base: 8 Nodes, Resolution: 1664 x 1664 x 832}} \\ \midrule
8              & 1.00                & 1.00                    \\ 
16             & 0.99                & 0.97                    \\ 
32             & 0.97                & 0.94                    \\ 
64             & 0.93                & 0.87                    \\  \midrule
\multicolumn{3}{c}{\textbf{Group 5, Base: 16 Nodes, Resolution: 1664 x 1664 x 1664}} \\ \midrule
16             & 1.00                & 1.00                    \\ 
32             & 0.93                & 0.95                    \\ 
64             & 0.91                & 0.94                    \\ 
128            & 0.89                & 0.87                    \\  \midrule
\multicolumn{3}{c}{\textbf{Group 6, Base: 32 Nodes, Resolution: 3328 x 1664 x 1664}} \\ \midrule
32             & 1.00                & 1.00                    \\ 
64             & 0.94                & 0.97                    \\ 
128            & 0.97                & 0.94                    \\ 
256            & 0.77                & 0.88                    \\ 
\bottomrule
\end{tabular}
\caption{Strong scaling Efficiencies at different resolutions for the tests presented in Fig.~\ref{fig:strong_nov} on MareNostrum 5 ACC and Leonardo Booster.}
\label{tab:scaling_efficiency_combined}
\end{table}
%




\subsection{3D Circularly Polarized Alfv\'en}

The propagation of circularly polarized Alfv\'en waves is a widely employed benchmark as it provides an exact nonlinear solution of the ideal MHD equations. 
Following \citep[][]{Mignone_2010}, we rotate a one-dimensional frame to obtain a 3D configuration and employ the same grid size used in Sec.~\ref{subsec:orszag_tang_test} for strong and weak scaling. 
The domain is a Cartesian cube with $L_x = L_y = L_z = 1$ with periodic boundaries.
The initial speed $\vec{v}$ and magnetic field $\vec{B}$ are chosen as
\begin{equation}
\label{eq:CPA_init}
  \begin{array}{l}
  \vec{v} = 
  \epsilon\sin(\phi) \hvec{e}_y + \epsilon \cos(\phi) \hvec{e}_y \;,
  \\ \noalign{\medskip}
  \vec{B} = \hvec{e}_x + \epsilon\sin(\phi) \hvec{e}_y + 
  \epsilon\cos(\phi) \hvec{e}_z \;,
   \\ \noalign{\medskip}
  \end{array}
\end{equation}
where $\epsilon = 0.1$ is the wave amplitude, while $\phi = \vec{k}\cdot\vec{x}$ with $\vec{k} = 2\pi (1/L_x,\, 1/L_y,\, 1/L_z)$.
Density and gas pressure are initially constant and equal to $\rho_0 = 1.0 $ and  $p_0 = 0.1$, respectively.
We employ an ideal equation of state with adiabatic index $\Gamma = 5/3$.

We solve the MHD equations using the $3^{\rm rd}$-order Runge-Kutta scheme with WENOZ reconstruction and the HLLD Riemann solver. 
The divergence-free condition is enforced using either divergence cleaning (for strong scaling) or constrained transport (for weak scaling) to differentiate the algorithms employed.

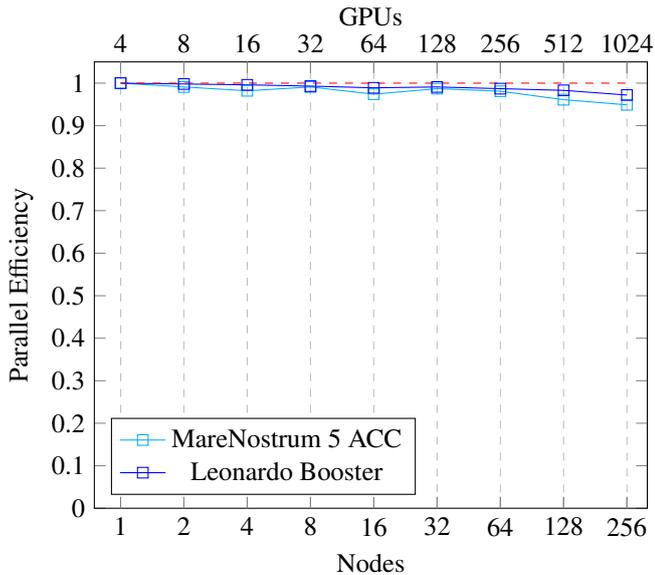
\begin{figure}[htbp]
\centering
\begin{tikzpicture}
\begin{axis}[
    width=\columnwidth,  
    height=7.5cm,  
    xlabel={Nodes},
    ylabel={Parallel Efficiency},
    xmin=0.75, xmax=320,
    ymin=0., ymax=1.05,
    xtick={1,2,4,8,16,32,64,128,256,512},
    xticklabels={1, 2, 4, 8, 16, 32, 64, 128, 256, 512},
    ytick={0.0,0.1,0.2,0.3,0.4,0.5,0.6,0.7,0.8,0.9,1.0},
    legend pos=south west,
    grid style=dashed,
    xmode=log,
    log basis x=2,
    extra x ticks={1,2,4,8,16,32,64,128,256,512},
    extra x tick labels={4,8,16,32,64,128,256,512,1024,2048}, 
    extra x tick style={grid=major, xticklabel pos=upper}, 
]
\addplot[color=cyan, mark=square]
    coordinates {
    (1,1.0)(2,0.991)(4,0.982)(8,0.991)(16,0.974)(32,0.987)(64,0.981)(128,0.961)(256,0.949)};
    \addlegendentry{MareNostrum 5 ACC}

\addplot[color=blue, mark=square]
    coordinates {
    (1,1.0)(2,0.998)(4,0.996)(8,0.993)(16,0.989)(32,0.991)(64,0.987)(128,0.983)(256,0.972)};
    \addlegendentry{Leonardo Booster}
\addplot[color=red, dashed]
    coordinates {
    (1,1.0)(2,1.0)(4,1.0)(8,1.0)(16,1.0)(32,1.0)(64,1.0)(128,1.0)(256,1.0)
    };
\end{axis}

\node[anchor=south, yshift=10pt] at (current axis.north) {GPUs};
\end{tikzpicture}
    \caption{Weak scaling tests of circularly polarized Alf\'ven waves on MareNostrum 5 ACC and Leonardo Booster.}
    \label{fig:weak_CPA}
\end{figure}

\begin{figure}[htbp]
\centering
\begin{tikzpicture}
\begin{axis}[
    width=\columnwidth,  
    height=7.5cm,  
    ylabel={Speedup},
    xlabel={Nodes},
    xmin=0, xmax=340,
    ymin=0, ymax=10,
    xtick={1,2,4,8,16,32,64,128,256},
    xticklabels={1, 2, 4, 8, 16, 32, 64, 128, 256},
    ytick={1,2,4,8},
    yticklabels={1, 2, 4, 8},
    legend pos=north west,
    legend style={at={(0.03,0.85)}}, 
    grid style=dashed,
    xmode=log,
    log basis x=2,
    ymode=log,
    log basis y=2,
    extra x ticks={1,2,4,8,16,32,64,128,256,512},
    extra x tick labels={4,8,16,32,64,128,256,512,1024,2048},
    extra x tick style={grid=major, xticklabel pos=upper},
]

\addplot[color=cyan, mark=square] coordinates {(1,1.0)(2,1.989)(4,3.87)(8,7.56)};
\addlegendentry{Marenostrum}

\addplot[color=blue, mark=square] coordinates {(1,1.0)(2,1.96)(4,3.87)(8,7.46)};
\addlegendentry{Leonardo}

\addplot[color=cyan, mark=square] coordinates {(2,1.0)(4,1.96)(8,3.72)(16,6.67)};
\addplot[color=cyan, mark=square] coordinates {(4,1.0)(8,1.97)(16,3.88)(32,6.21)};
\addplot[color=cyan, mark=square] coordinates {(8,1.0)(16,2.0)(32,3.75)(64,6.73)};
\addplot[color=cyan, mark=square] coordinates {(16,1.0)(32,1.94)(64,3.83)(128,6.18)};
\addplot[color=cyan, mark=square] coordinates {(32,1.0)(64,1.93)(128,3.47)(256,6.92)};

\addplot[color=blue, mark=square] coordinates {(2,1.0)(4,1.94)(8,3.87)(16,7.68)};
\addplot[color=blue, mark=square] coordinates {(4,1.0)(8,1.96)(16,3.89)(32,7.57)};
\addplot[color=blue, mark=square] coordinates {(8,1.0)(16,1.97)(32,3.81)(64,7.63)};
\addplot[color=blue, mark=square] coordinates {(16,1.0)(32,1.95)(64,3.89)(128,7.66)};
\addplot[color=blue, mark=square] coordinates {(32,1.0)(64,1.97)(128,3.90)(256,7.50)};

\addplot[color=red, dashed] coordinates {(0,0)(1,1.0)(2,2)(4,4)(8,8)(16,16)};
\addplot[color=red, dashed] coordinates {(2,1)(4,2)(8,4)(16,8)(32,16)};
\addplot[color=red, dashed] coordinates {(4,1)(8,2)(16,4)(32,8)(64,16)};
\addplot[color=red, dashed] coordinates {(8,1)(16,2)(32,4)(64,8)(128,16)};
\addplot[color=red, dashed] coordinates {(16,1)(32,2)(64,4)(128,8)(256,16)};
\addplot[color=red, dashed] coordinates {(32,1)(64,2)(128,4)(256,8)(512,16)};
\end{axis}
\node[anchor=south, yshift=10pt] at (current axis.north) {GPUs};
\end{tikzpicture}
\caption{Strong Scaling tests of Circularly Polarized Alf\'ven on MareNostrum 5 ACC and Leonardo Booster.}
    \label{fig:strong_CPA}
\end{figure}
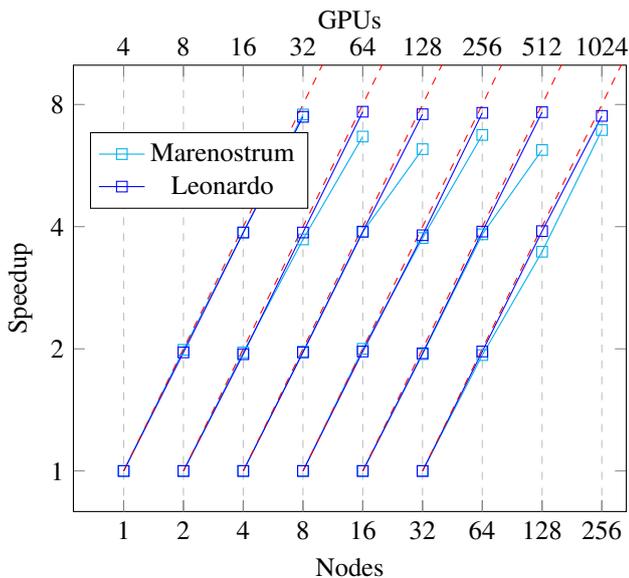

All computations are performed in double-precision arithmetic.
The tests have been performed on Leonardo Booster and MareNostrum 5 ACC partitions, using the same grid resolution (Tab.~\ref{tab:weak_nov}) of \ref{subsec:orszag_tang_test},  

Fig.~\ref{fig:weak_CPA} presents the parallel efficiency expressed as the ratio of the wall clock times on $1$ node and $N$ nodes, $T_{1}/T_{N}$.
The efficiency decreases slightly as the number of nodes grows due to the increase in communications and physical distances between the devices. 
The best results are obtained, this time, with Leonardo with an efficiency of $0.97$ for 256 nodes (1024 GPUs), while we obtain $0.95$ on MareNostrum 5. 
Strong scaling test results are presented in Tab.~\ref{tab:scaling_efficiency_combined_CPA} and in Fig.~\ref{fig:strong_CPA}. 
As for the Orszag-Tang test, six strong scaling tests are presented for four groups of runs with fixed resolution. 
As expected, the strong scaling efficiency slightly decreases as the number of nodes rises. 
Leonardo Booster's results are marginally better than MareNostrum 5 ACC for this test, with an average efficiency of $0.97$ and $0.93$, respectively. 

\begin{table}
\centering
\begin{tabular}{ccc}
\toprule
\textbf{Nodes} & \textbf{MareNostrum 5 Efficiency} & \textbf{Leonardo Efficiency} \\ \midrule
\multicolumn{3}{c}{\textbf{Group 1, Base: 1 Node, Resolution: 832 x 832 x 416}} \\ \midrule
1              & 1.00                & 1.00                    \\ 
2              & 0.98                & 0.97                    \\ 
4              & 0.93                & 0.97                    \\ 
8              & 0.83                & 0.96                    \\ \midrule
\multicolumn{3}{c}{\textbf{Group 2, Base: 2 Nodes, Resolution: 832 x 832 x 832}} \\ \midrule
2              & 1.00                & 1.00                    \\ 
4              & 0.98                & 0.98                    \\ 
8              & 0.97                & 0.98                    \\ 
16             & 0.78                & 0.95                    \\ \midrule
\multicolumn{3}{c}{\textbf{Group 3, Base: 4 Nodes, Resolution: 1664 x 832 x 832}} \\ \midrule
4              & 1.00                & 1.00                    \\ 
8              & 1.00                & 0.99                    \\ 
16             & 0.94                & 0.95                    \\ 
32             & 0.84                & 0.95                    \\ \midrule
\multicolumn{3}{c}{\textbf{Group 4, Base: 8 Nodes, Resolution: 1664 x 1664 x 832}} \\ \midrule
8              & 1.00                & 1.00                    \\ 
16             & 0.97                & 0.98                    \\ 
32             & 0.95                & 0.97                    \\ 
64             & 0.81                & 0.96                    \\ \midrule
\multicolumn{3}{c}{\textbf{Group 5, Base: 16 Nodes, Resolution: 1664 x 1664 x 1664}} \\ \midrule
16             & 1.00                & 1.00                    \\ 
32             & 0.97                & 0.99                    \\ 
64             & 0.96                & 0.97                    \\ 
128            & 0.77                & 0.94                    \\ \midrule
\multicolumn{3}{c}{\textbf{Group 6, Base: 32 Nodes, Resolution: 3328 x 1664 x 1664}} \\ \midrule
32             & 1.00                & 1.00                    \\ 
64             & 0.96                & 0.99                    \\ 
128            & 0.87                & 0.97                    \\ 
256            & 0.87                & 0.94                    \\ \bottomrule
\end{tabular}
\caption{Strong scaling Efficiencies at different resolution for the tests presented in Fig. \ref{fig:strong_CPA} on
MareNostrum 5 ACC and Leonardo Booster.}
\label{tab:scaling_efficiency_combined_CPA}
\end{table}

\subsection{Energy Efficiency}

Energy efficiency for an application is defined as performance in Floating point Operations Per Second (FLOPS) per unit of power consumption (Watt). 
The total energy consumed (in Joules) is approximated from power samples (in Watts) obtained at a specific sampling frequency (in Hertz) from a power monitoring system, as depicted in the following equation:

\begin{equation} \label{eq:energy}
\mathrm{Energy}(t) = \int_0^t \mathrm{Power}(t')\mathrm{d}t' \approx \frac{\sum_{i=0}^n P_i}{f_s},\nonumber
\end{equation}
where $Ps_i$ is i-th power sample, $f_s$ is sampling frequency. 

\subsubsection{Methodology}
For energy measurements, we used GPU-accelerated nodes of the EuroHPC Karolina cluster~\citep[][]{Karolina} equipped with two AMD EPYC 7763 CPUs and eight NVIDIA A100-SXM4 GPUs. 
The energy consumption of the GPU is measured using performance counters of the GPU via NVIDIA Management Library (NVML)~\citep[][]{Nvidia-NVML}. 
For CPU, we use AMD RAPL~\citep[][]{AMDrapl,Zen2analysis}. 
The power consumption of the remaining components of the compute node is estimated using a simple mathematical model
(see Tab.~\ref{tab:EE-hw-platform-consumption}). 

To investigate the energy efficiency of the application, we have tuned the frequency of the GPUs' Streaming Multiprocessors (SMs), which is analogous to the CPU core frequency. 
For this purpose, we use NVML function 
\Cinline{nvmlDeviceSetApplicationsClocks()} that sets a specific clock speed of a GPU SMs. 
Since A100-SXM4 uses HBM2 memory, its frequency cannot be tuned; 
on GPUs with GDDR memory, on the other hand, the
frequency can be scaled using the same function. 

We use static frequency tuning, meaning that a single GPU SMs frequency is applied at the start of the application execution and remains constant until the end. 
The key advantage of static tuning is that it can be performed by a standard HPC job scheduler like SLURM~\citep[][]{SLURM} or PBS~\citep[][]{PBSPro} at the job launch time. 
The disadvantage of this method is that the applied configuration may not be optimal for all executed kernels. 
Thus, dynamic tuning may bring more energy savings for the same, or even less, performance penalty~\citep[][]{READEXbook}.

\subsubsection{Platform analysis}
\begin{table}[htbp]
    \centering
    \resizebox{\linewidth}{!}{%
    \small
    \begin{tabular}{
        >{\raggedright\arraybackslash}p{0.21\linewidth}
        >{\raggedright\arraybackslash}p{0.16\linewidth}
        >{\raggedright\arraybackslash}p{0.14\linewidth}
        >{\raggedright\arraybackslash}p{0.17\linewidth}
        >{\raggedright\arraybackslash}p{0.11\linewidth}
    }
    \toprule
    \textbf{Karolina GPU node} & \textbf{CPUs} & \textbf{GPUs} & \textbf{Other parts} & \textbf{Total} \\
    \midrule
    Units per server & 2× AMD EPYC 7763 & 8× NVIDIA A100 & -- & -- \\\midrule
    Range of power consumption & 90–280\,W per CPU & 55–400\,W per GPU & 520–625\,W per server & -- \\\midrule
    Static power consumption   & 180\,W & 420\,W & 520\,W & 1120\,W \\\midrule
    Dynamic power consumption  & max. 380\,W & max. 2780\,W & max. 105\,W & max. 3265\,W \\
    \midrule \midrule
    \pluto{} power consumption range & 115–120\,W per CPU & 99–248\,W per GPU & 600–610\,W & 1622– 2834\,W \\ \midrule
    \pluto{} dynamic pow. consumption & 5\,W per CPU; 10\,W total & 149\,W per GPU; 1192\,W total & 10\,W & 1212\,W \\
    \bottomrule
    \end{tabular}
    }
    \caption{Power consumption of the key components of the GPU accelerated compute node of the Karolina supercomputer. Static power consumption is always present, even when a server is idle. Dynamic power consumption is additional power consumption when a server is under load (CPUs and GPUs perform some computations) and it is workload-dependent. The large \gpluto{} power consumption range for GPUs is caused by both lowering the SM frequency to 990\,GHz as well as lowering GPU utilization for the high number of compute nodes in the strong scaling test.}
    \label{tab:EE-hw-platform-consumption}
\end{table}
The key properties of the Karolina GPU-accelerated node from the energy consumption point of view are shown in Tab.~\ref{tab:EE-hw-platform-consumption}. 
We break the hardware into 3 key groups: CPUs, GPUs, and the rest of the server (including motherboard, memory DIMMs, network cards, GPU board with NVlink switches, etc.). 
For each group, we show the "Range of power consumption", where the lower number is the power consumption of the component when the server is idle (no computations are running) and the higher value presents maximum power consumption under full load. 
When all values for idle power consumption for all components are summed, we get the "Static power consumption" of the entire server. 
In this case, just to keep the server turned on, it consumes 1.12\,kW. 
The "Dynamic power consumption" then denotes the additional power that can be consumed when the server is under full load. This is a maximum/theoretical value if a workload is capable of full utilization of the CPUs or GPUs.  
By summing the static and dynamic consumption, we obtain the server's peak power consumption. 
Our methodology can only control and reduce dynamic power consumption. 
From the server user perspective, the static power cannot be controlled, and it is the main reason why energy consumption increases when runtime is extended, as energy is the integral of power over time.  
Finally, the last two rows in the Tab. present the platform's behavior when \gpluto{} runs the 3D Orszag-Tang vortex benchmark. 
We can see that CPUs and remaining server components consumption is mostly static (because we do not tune CPU frequency), 
while the key component with predominantly dynamic power consumption are the 8 GPUs.
In summary, each compute node can consume between 1.6 and 2.8\,kW depending on how many nodes are used for the experiment (we use strong scaling) and the selected frequency of the GPU SMs, resulting in 1.2\,kW dynamically controllable power consumption.

\begin{figure}[htbp]
    \centering    
    \includegraphics[width=\columnwidth]{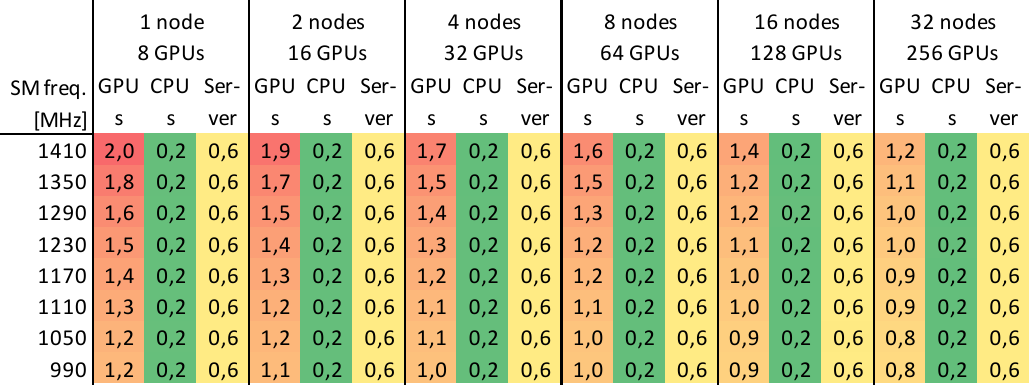}
    
    \hspace{1mm}
    
    \includegraphics[width=\columnwidth]{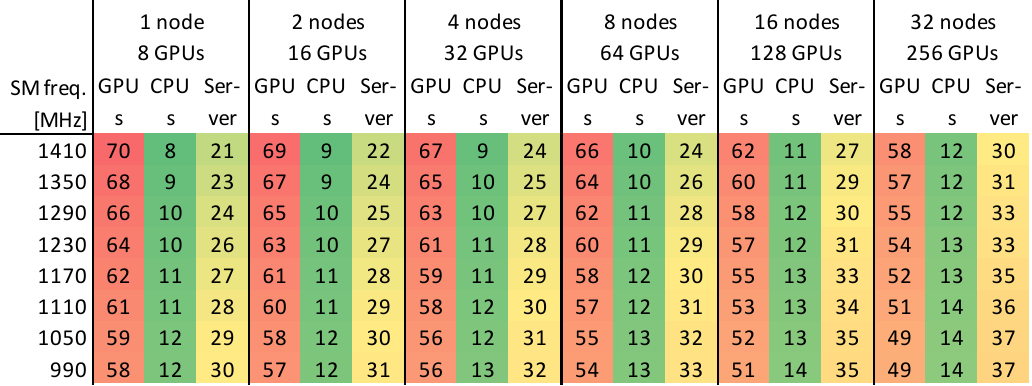}
    
    \caption{Power consumption per node for the 3D Orszag-Tang Vortex benchmark. 
    The top panel shows absolute values in [kW].
    The bottom one shows the relative values with respect to the entire node power consumption in [\%]. 
    }
    
    \label{fig:EE-Orsag-power}
\end{figure}

The compute node behavior is presented in Fig.~\ref{fig:EE-Orsag-power} in more detail. 
We can see that: 
(a) by changing the SM frequency from the default 1.41\,GHz to 0.99\,GHz the power consumption of all 8 GPUs per node is reduced from 2.0 to 1.2\,kW (when the benchmark runs on one node); 
(b) by performing the strong scaling from 1 to 32 compute nodes, we also reduce GPUs power consumption per node from 2.0 to 1.2\,kW. This is caused by the strong scaling because the problem size per GPU is halved with every step of strong scaling. This results in lower utilization of the GPUs and, therefore, lower average power consumption.     
For 32 nodes, if we scale the SM frequency down to 0.99\,GHz, we can further reduce GPU power consumption to 0.8\,kW. 
It is important to realize that,
for this configuration, the static power consumption of the non-GPU components amounts to half the total.
When this effect is combined with the extended runtime due to SM frequency reduction (for 32 nodes, reducing the SM frequency from 1.4 to 0.99\,GHz yields a runtime extension to $117.6\,\%$ - see Figure~\ref{fig:EE-Orsag-energy} top) it becomes dominant in terms of energy consumption of the computation. 
This explains why the highest energy savings (see bottom panels of Fig.~\ref{fig:EE-Orsag-energy} and \ref{fig:EE-CPAlfven-energy}) are not achieved for the lowest SM frequency we used.

\subsubsection{Energy consumption optimization results}

\begin{figure}[htbp]
    \centering    
    \includegraphics[width=\columnwidth]{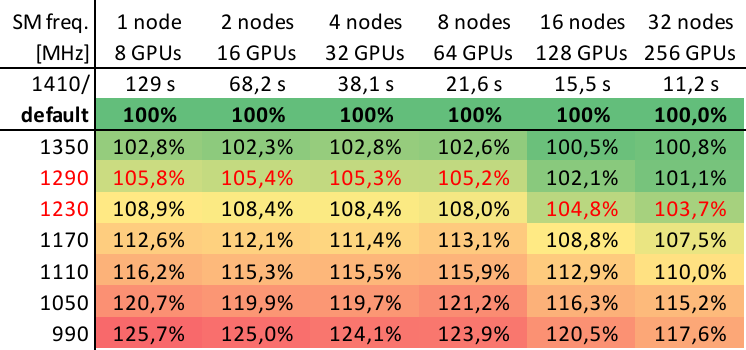}

    \hspace{1mm}

    \includegraphics[width=\columnwidth]{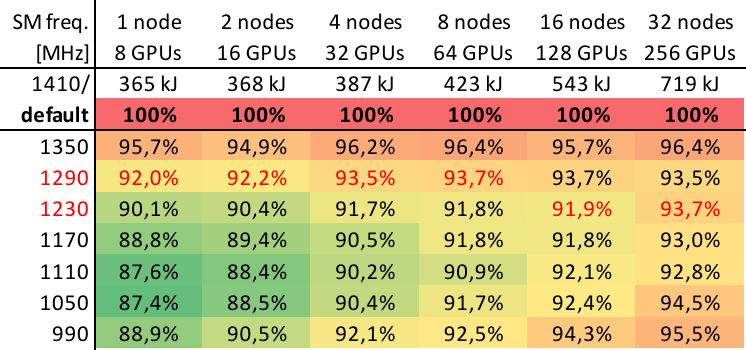}

    \caption{Impact of the static tuning of the GPU SM frequency on the runtime (top) and energy consumption (bottom) of the 3D Orszag-Tang vortex benchmark. 
    The top panel shows runtime variations with respect to the default execution time shown in the first line.
    The bottom one shows relative energy consumption with respect to the energy consumption of the default execution, as shown in the first line of the table.
    }
    \label{fig:EE-Orsag-energy}
\end{figure}

\begin{figure}[htbp]
    \centering    
    \includegraphics[width=\columnwidth]{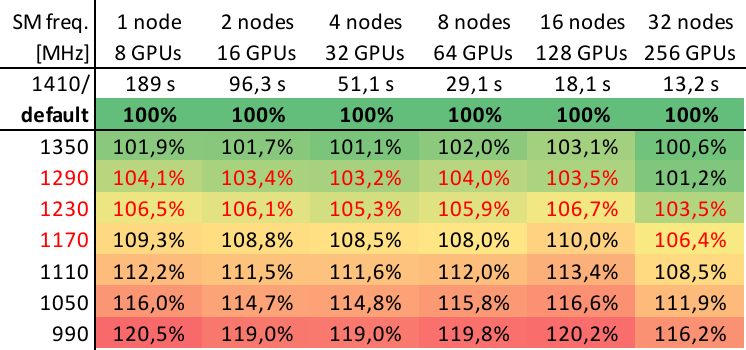}

    \hspace{1mm}

    \centering    
    \includegraphics[width=\columnwidth]{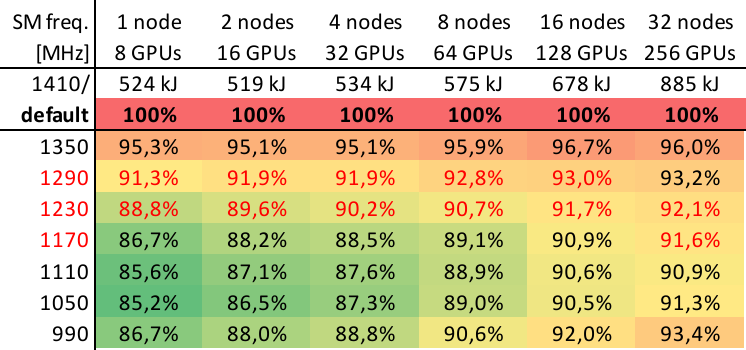}

    \caption{Same as Fig. \ref{fig:EE-Orsag-energy} but for the 3D circularly polarized Alfv\'en test.
    }
    \label{fig:EE-CPAlfven-energy}
\end{figure}
We have performed strong scaling from 1 to 32 GPU-accelerated compute nodes of Karolina for both  
3D circularly polarized Alfv\'en wave 
and 
3D Orszag-Tang vortex 
benchmarks. 
For each number of nodes, we scaled down the SM frequency from 1.4 to 0.99\,GHz and measured the runtime and energy consumption of all compute nodes employed for the benchmark. 
The measurement results are in Fig.~\ref{fig:EE-Orsag-energy} and \ref{fig:EE-CPAlfven-energy}. 
We do not present results for frequencies lower than 0.99\,GHz because, for all cases, the energy savings for 0.99\,GHz are smaller than the savings for 1.11 and 1.05\,GHz. 

If we consider acceptable an increase in runtime up to 6\% (values close to this threshold are highlighted in red in both figures), we see that we can save
up to $8\,\%$ of energy for the 3D Orszag-Tang vortex and up to $11\,\%$ for the 3D circularly polarized Alfv\'en wave benchmark. 
We also see that 
by increasing the number of compute nodes 
we gain lower savings.
As mentioned previously, this is due to a high ratio of static power consumption and extended runtime. 

We want to highlight that these servers are air-cooled, and therefore, the energy consumption of the remaining components is relatively high. By using modern hot water-cooled servers and racks, the static power consumption 
can be reduced, with a corresponding increase in energy savings.

\section{Summary}
\label{sec:summary}
%
%

In this paper we have presented a first look into \gpluto{}, a GPU-optimized implementation of the widely used \pluto{} code for computational plasma astrophysics. 
The new code is written in C++ and adopts OpenACC as a programming model to efficiently exploit modern GPU architectures. 
The new implementation maintains the flexibility and robustness of the legacy code while significantly improving computational performance, power efficiency, and scalability on HPC clusters.

This paper outlines the fundamental design principles behind \gpluto{}, including the numerical methods employed, the OpenACC-based parallelization strategy, memory management optimizations, and multi-GPU communication. 
The performance of \gpluto{} has been evaluated through two numerical benchmarks: the 3D Orszag-Tang vortex and the 3D circularly polarized Alfv{\'e}n wave, conducted on three European HPC platforms: Marenostrum 5, Leonardo, and Meluxina. 
Both weak and strong scaling tests have been conducted.

Results demonstrate excellent strong and weak scaling efficiencies, maintaining around $90\%$ performance across multiple nodes, in a situation of high GPU occupancy.
Direct comparison between the GPU- and CPU-specialized partitions of Leonardo reveals a factor $\approx 9$ acceleration (per node usage).
The results of this study confirm that \gpluto{} successfully harnesses the power of GPUs to enhance the performance of MHD simulations. 

The strong scaling behavior suggests that \gpluto{} can efficiently utilize a large number of GPUs, making it well-suited for exascale computing applications. 

Energy efficiency analyses highlight the impact of GPU frequency tuning in optimizing computational power consumption. 
Our experiments indicate that adjusting GPU clock frequencies can lead to significant energy savings with minimal runtime penalties, highlighting potential avenues for optimizing HPC resource utilization. 
In particular, lowering the SMs frequency can save around $8\,\%$ and $11\,\%$ of energy consumption for 3D Orszag-Tang tests and circularly polarized Alfvén waves, respectively, at a cost of $6\,\%$ longer execution time.

The development of \gpluto{} represents a significant step toward enabling more efficient and scalable astrophysical simulations on next-generation, Exascale supercomputers.
A more comprehensive description of the code along with all its physical modules (including relativistic and non-ideal MHD extensions, particles and different systems of coordinates) and the implemented numerical methods - including high-order methods \citep{berta2024,mignone2024} - will be presented in a forthcoming work. 
We also invite to refer to the companion paper \citep{suriano2025} for the Lagrangian Particles module integration in gPLUTO.
Future works on HPC porting will also explore additional acceleration frameworks such as OpenMP for execution on AMD-based HPC systems.


\section*{Acknowledgements}

MR and AM wish to thank F. Spiga from NVIDIA for his active support during the very early stages of this development. 
MR and AM wish also to thank M. Bettencourt for his technical guidance and precious advices while porting the code to C++.

\vspace{7pt}\noindent
This work has received funding from the European High Performance Computing Joint Undertaking (JU) and Belgium, Czech Republic, France, Germany, Greece, Italy, Norway, and Spain under grant agreement No 101093441 (SPACE).

\vspace{7pt}\noindent
This project has received funding from the European Union's Horizon Europe research and innovation programme under the Marie Sk\l{}odowska-Curie grant agreement No. 101064953 (GR-PLUTO).

\vspace{7pt}\noindent
This work was supported by the Ministry of Education, Youth and Sports of the Czech Republic through the e-INFRA CZ (ID:90254). 

\vspace{7pt}\noindent
We acknowledge ISCRA for awarding this project access to the Leonardo supercomputer and the EuroHPC Joint Undertaking for granting us access to the Leonardo, MareNostrum and MeluXina supercomputer through an EuroHPC [Extreme/Regular/Benchmark/Development/…] Access call.

\vspace{7pt}\noindent
This paper is supported by the Fondazione ICSC, Spoke 3 Astrophysics and Cosmos Observations and National Recovery and Resilience Plan (Piano Nazionale di Ripresa e Resilienza, PNRR), Project ID CN\_00000013 “Italian Research Center on High-Performance Computing, Big Data and Quantum Computing” funded by MUR Missione 4 Componente 2 Investimento 1.4: Potenziamento strutture di ricerca e creazione di “campioni nazionali di R\&S (M4C2-19)” - Next Generation EU (NGEU).

\bibliographystyle{elsarticle-num}
\bibliography{paper,meric,energy-bibtex} 

@Misc{Nvidia-NVML,
author =   {{NVIDIA Corp.}},
title =    {{NVML} {API Reference}},
howpublished = {\url{https://docs.nvidia.com/deploy/nvml-api/nvml-api-reference.html}},
year = {2025},
month = {January},
note = {"[Online; accessed 2025-02-28]"}
}

@book{READEXbook,
  title     = {System-Scenario-based Design Principles and Applications},
  author    = {Catthoor, Francky and Basten, Twan and Zompakis, Nikolaos and Geilen, Marc and Kjeldsberg, Per Gunnar},
  publisher = {Springer International Publishing},
  year      = {2020},
  series    = 1,
  isbn = "978-3-030-20342-9",
  doi = "10.1007/978-3-030-20343-6"
}

@article{Zen2analysis,
  author       = {Schone, Robert and
                  Ilsche, Thomas and
                  Bielert, Mario and
                  Velten, Markus and
                  Schmidl, Markus and
                  Hackenberg, Daniel },
  title        = {Energy Efficiency Aspects of the {AMD} Zen 2 Architecture},
  journal      = {2021 IEEE International Conference on Cluster Computing (CLUSTER)},
  volume       = {abs/2108.00808},
  year         = {2021},
 pages = {562-571},
  url          = {https://arxiv.org/abs/2108.00808}
}

@misc{AMDrapl,
  author = {{AMD}},
  title = {Kernel driver amd\_energy},
  howpublished = {\url{https://github.com/amd/amd_energy}},
  note = {"[Online; accessed 2025-02-28]"}
}

@Misc{Karolina,
author =   {{IT4Innovations}},
title =    {{Karolina supercomputer} {User Guide}},
howpublished = {\url{https://docs.it4i.cz/karolina/introduction/}},
year = {2025},
month = {February},
note = {"[Online; accessed 2025-02-28]"}
}

@inbook{PBSPro,
author = {Nitzberg, Bill and Schopf, Jennifer M. and Jones, James Patton},
title = {PBS Pro: Grid computing and scheduling attributes},
year = {2004},
isbn = {1402075758},
publisher = {Kluwer Academic Publishers},
address = {USA},
booktitle = {Grid Resource Management: State of the Art and Future Trends},
pages = {183–190},
numpages = {8}
}

@InProceedings{SLURM,
author="Yoo, Andy B.
and Jette, Morris A.
and Grondona, Mark",
editor="Feitelson, Dror
and Rudolph, Larry
and Schwiegelshohn, Uwe",
title="SLURM: Simple Linux Utility for Resource Management",
booktitle="Job Scheduling Strategies for Parallel Processing",
year="2003",
publisher="Springer Berlin Heidelberg",
address="Berlin, Heidelberg",
pages="44--60",
isbn="978-3-540-39727-4"
}

@ARTICLE{suriano2025,
       author = {{Suriano}, A. and {Truzzi}, S. and {Costa}, A. and {Rossazza}, M. and {Shukla}, N. and {Berta}, V. and {Mignone}, A. and {Zanni}, C},
       title = "{The PLUTO Code on GPUs: Offloading Lagrangian Particle Methods}",
       year=2025,
      journal = {Manuscript submitted for publication (companion paper)},
}

@ARTICLE{amato2018,
       author = {{Amato}, Elena and {Blasi}, Pasquale},
        title = "{Cosmic ray transport in the Galaxy: A review}",
      journal = {Advances in Space Research},
         year = 2018,
        month = nov,
       volume = {62},
       number = {10},
        pages = {2731-2749},
          doi = {10.1016/j.asr.2017.04.019},
archivePrefix = {arXiv},
       eprint = {1704.05696},
 primaryClass = {astro-ph.HE},
       adsurl = {https://ui.adsabs.harvard.edu/abs/2018AdSpR..62.2731A}
}

@ARTICLE{Balsara_Spicer1999,
   author = {{Balsara}, D.~S. and {Spicer}, D.~S.},
    title = "{A Staggered Mesh Algorithm Using High Order Godunov Fluxes to Ensure Solenoidal Magnetic Fields in Magnetohydrodynamic Simulations}",
  journal = {Journal of Computational Physics},
     year = 1999,
    month = mar,
   volume = 149,
    pages = {270-292},
      doi = {10.1006/jcph.1998.6153},
   adsurl = {http://adsabs.harvard.edu/abs/1999JCoPh.149..270B}
}

@ARTICLE{berta2024,
       author = {{Berta}, V. and {Mignone}, A. and {Bugli}, M. and {Mattia}, G.},
        title = "{A 4$^{th}$-order accurate finite volume method for ideal classical and special relativistic MHD based on pointwise reconstructions}",
      journal = {Journal of Computational Physics},
         year = 2024,
        month = feb,
       volume = {499},
          eid = {112701},
        pages = {112701},
          doi = {10.1016/j.jcp.2023.112701},
archivePrefix = {arXiv},
       eprint = {2310.11831},
 primaryClass = {astro-ph.HE},
       adsurl = {https://ui.adsabs.harvard.edu/abs/2024JCoPh.49912701B}
}

@ARTICLE{Borges_WENOZ2008,
       author = {{Borges}, Rafael and {Carmona}, Monique and {Costa}, Bruno and {Don}, Wai Sun},
        title = "{An improved weighted essentially non-oscillatory scheme for hyperbolic conservation laws}",
      journal = {Journal of Computational Physics},
         year = 2008,
        month = mar,
       volume = {227},
       number = {6},
        pages = {3191-3211},
          doi = {10.1016/j.jcp.2007.11.038},
       adsurl = {https://ui.adsabs.harvard.edu/abs/2008JCoPh.227.3191B}
}

@ARTICLE{cattorini2024,
       author = {{Cattorini}, Federico and {Giacomazzo}, Bruno},
        title = "{GRMHD study of accreting massive black hole binaries in astrophysical environment: A review}",
      journal = {Astroparticle Physics},
         year = 2024,
        month = jan,
       volume = {154},
          eid = {102892},
        pages = {102892},
          doi = {10.1016/j.astropartphys.2023.102892},
archivePrefix = {arXiv},
       eprint = {2401.02521},
 primaryClass = {astro-ph.HE},
       adsurl = {https://ui.adsabs.harvard.edu/abs/2024APh...15402892C}
}

@ARTICLE{charbonneau2020,
       author = {{Charbonneau}, Paul},
        title = "{Dynamo models of the solar cycle}",
      journal = {Living Reviews in Solar Physics},
         year = 2020,
        month = dec,
       volume = {17},
       number = {1},
          eid = {4},
        pages = {4},
          doi = {10.1007/s41116-020-00025-6},
       adsurl = {https://ui.adsabs.harvard.edu/abs/2020LRSP...17....4C}
}

@ARTICLE{Colella_Woodward1984,
       author = {{Colella}, P. and {Woodward}, Paul R.},
        title = "{The Piecewise Parabolic Method (PPM) for Gas-Dynamical Simulations}",
      journal = {Journal of Computational Physics},
         year = 1984,
        month = sep,
       volume = {54},
        pages = {174-201},
          doi = {10.1016/0021-9991(84)90143-8},
       adsurl = {https://ui.adsabs.harvard.edu/abs/1984JCoPh..54..174C}
}

@ARTICLE{Dedner_etal2002,
   author = {{Dedner}, A. and {Kemm}, F. and {Kr{\"o}ner}, D. and {Munz}, C.-D. and 
	{Schnitzer}, T. and {Wesenberg}, M.},
    title = "{Hyperbolic Divergence Cleaning for the MHD Equations}",
  journal = {Journal of Computational Physics},
     year = 2002,
    month = jan,
   volume = 175,
    pages = {645-673},
      doi = {10.1006/jcph.2001.6961},
   adsurl = {http://adsabs.harvard.edu/abs/2002JCoPh.175..645D}
}

@ARTICLE{delzanna2022,
       author = {{Del Zanna}, Luca and {Tomei}, Niccol{\`o} and {Franceschetti}, Kevin and {Bugli}, Matteo and {Bucciantini}, Niccol{\`o}},
        title = "{General Relativistic Magnetohydrodynamics Mean-Field Dynamos}",
      journal = {Fluidika},
         year = 2022,
        month = feb,
       volume = {7},
       number = {2},
        pages = {87},
          doi = {10.3390/fluids7020087},
       adsurl = {https://ui.adsabs.harvard.edu/abs/2022Fluid...7...87D}
}

@article{delzanna2024,
  title = {A {{GPU-Accelerated Modern Fortran Version}} of the {{ECHO Code}} for {{Relativistic Magnetohydrodynamics}}},
  author = {Del Zanna, Luca and Landi, Simone and Serafini, Lorenzo and Bugli, Matteo and Papini, Emanuele},
  year = {2024},
  month = jan,
  journal = {Fluids},
  volume = {9},
  number = {1},
  pages = {16},
  publisher = {Multidisciplinary Digital Publishing Institute},
  issn = {2311-5521},
  doi = {10.3390/fluids9010016},
  urldate = {2024-01-06},
  copyright = {http://creativecommons.org/licenses/by/3.0/},
  langid = {english}
}

@ARTICLE{doyle2007,
       author = {{Doyle}, E.~J. and {Houlberg}, W.~A. and {Kamada}, Y. and {Mukhovatov}, V. and {Osborne}, T.~H. and {Polevoi}, A. and {Bateman}, G. and {Connor}, J.~W. and {Cordey}, J.~G. and {Fujita}, T. and {Garbet}, X. and {Hahm}, T.~S. and {Horton}, L.~D. and {Hubbard}, A.~E. and {Imbeaux}, F. and {Jenko}, F. and {Kinsey}, J.~E. and {Kishimoto}, Y. and {Li}, J. and {Luce}, T.~C. and {Martin}, Y. and {Ossipenko}, M. and {Parail}, V. and {Peeters}, A. and {Rhodes}, T.~L. and {Rice}, J.~E. and {Roach}, C.~M. and {Rozhansky}, V. and {Ryter}, F. and {Saibene}, G. and {Sartori}, R. and {Sips}, A.~C.~C. and {Snipes}, J.~A. and {Sugihara}, M. and {Synakowski}, E.~J. and {Takenaga}, H. and {Takizuka}, T. and {Thomsen}, K. and {Wade}, M.~R. and {Wilson}, H.~R. and {ITPA Transport Physics Topical Group} and {Confinement Database}, ITPA and {Modelling Topical Group} and {Pedestal}, ITPA and {Edge Topical Group}},
        title = "{Chapter 2: Plasma confinement and transport}",
      journal = {Nuclear Fusion},
         year = 2007,
        month = jun,
       volume = {47},
       number = {6},
        pages = {S18-S127},
          doi = {10.1088/0029-5515/47/6/S02},
       adsurl = {https://ui.adsabs.harvard.edu/abs/2007NucFu..47S..18D}
}

@ARTICLE{Evans_Hawley1988,
   author = {{Evans}, C.~R. and {Hawley}, J.~F.},
    title = "{Simulation of magnetohydrodynamic flows - A constrained transport method}",
  journal = {The Astrophysical Journal},
     year = 1988,
    month = sep,
   volume = 332,
    pages = {659-677},
      doi = {10.1086/166684},
   adsurl = {http://adsabs.harvard.edu/abs/1988ApJ...332..659E}
}

@ARTICLE{faber2012,
       author = {{Faber}, Joshua A. and {Rasio}, Frederic A.},
        title = "{Binary Neutron Star Mergers}",
      journal = {Living Reviews in Relativity},
         year = 2012,
        month = dec,
       volume = {15},
       number = {1},
          eid = {8},
        pages = {8},
          doi = {10.12942/lrr-2012-8},
archivePrefix = {arXiv},
       eprint = {1204.3858},
 primaryClass = {gr-qc},
       adsurl = {https://ui.adsabs.harvard.edu/abs/2012LRR....15....8F}
}

@ARTICLE{Helzel_etal2011,
       author = {{Helzel}, Christiane and {Rossmanith}, James A. and {Taetz}, Bertram},
        title = "{An unstaggered constrained transport method for the 3D ideal magnetohydrodynamic equations}",
      journal = {Journal of Computational Physics},
         year = "2011",
        month = "May",
       volume = {230},
       number = {10},
        pages = {3803-3829},
          doi = {10.1016/j.jcp.2011.02.009},
archivePrefix = {arXiv},
       eprint = {1007.2606},
 primaryClass = {math.NA},
       adsurl = {https://ui.adsabs.harvard.edu/abs/2011JCoPh.230.3803H}
}

@ARTICLE{inghirami2016,
       author = {{Inghirami}, Gabriele and {Del Zanna}, Luca and {Beraudo}, Andrea and {Moghaddam}, Mohsen Haddadi and {Becattini}, Francesco and {Bleicher}, Marcus},
        title = "{Numerical magneto-hydrodynamics for relativistic nuclear collisions}",
      journal = {European Physical Journal C},
         year = 2016,
        month = dec,
       volume = {76},
       number = {12},
          eid = {659},
        pages = {659},
          doi = {10.1140/epjc/s10052-016-4516-8},
archivePrefix = {arXiv},
       eprint = {1609.03042},
 primaryClass = {hep-ph},
       adsurl = {https://ui.adsabs.harvard.edu/abs/2016EPJC...76..659I}
}

@ARTICLE{janka2025,
       author = {{Janka}, H. -Thomas},
        title = "{Long-Term Multidimensional Models of Core-Collapse Supernovae: Progress and Challenges}",
      journal = {arXiv e-prints},
         year = 2025,
        month = feb,
          eid = {arXiv:2502.14836},
        pages = {arXiv:2502.14836},
          doi = {10.48550/arXiv.2502.14836},
archivePrefix = {arXiv},
       eprint = {2502.14836},
 primaryClass = {astro-ph.HE},
       adsurl = {https://ui.adsabs.harvard.edu/abs/2025arXiv250214836J}
}

@ARTICLE{kalinani2025,
       author = {{Kalinani}, Jay V. and {Ji}, Liwei and {Ennoggi}, Lorenzo and {Lopez Armengol}, Federico G. and {Timotheo Sanches}, Lucas and {Tsao}, Bing-Jyun and {Brandt}, Steven R. and {Campanelli}, Manuela and {Ciolfi}, Riccardo and {Giacomazzo}, Bruno and {Haas}, Roland and {Schnetter}, Erik and {Zlochower}, Yosef},
        title = "{AsterX: a new open-source GPU-accelerated GRMHD code for dynamical spacetimes}",
      journal = {Classical and Quantum Gravity},
         year = 2025,
        month = jan,
       volume = {42},
       number = {2},
          eid = {025016},
        pages = {025016},
          doi = {10.1088/1361-6382/ad9c11},
archivePrefix = {arXiv},
       eprint = {2406.11669},
 primaryClass = {astro-ph.HE},
       adsurl = {https://ui.adsabs.harvard.edu/abs/2025CQGra..42b5016K}
}

@ARTICLE{knaepen2008,
       author = {{Knaepen}, Bernard and {Moreau}, Ren{\'e}},
        title = "{Magnetohydrodynamic Turbulence at Low Magnetic Reynolds Number}",
      journal = {Annual Review of Fluid Mechanics},
         year = 2008,
        month = jan,
       volume = {40},
       number = {1},
        pages = {25-45},
          doi = {10.1146/annurev.fluid.39.050905.110231},
       adsurl = {https://ui.adsabs.harvard.edu/abs/2008AnRFM..40...25K}
}

@ARTICLE{lazarian2012,
       author = {{Lazarian}, A. and {Vlahos}, L. and {Kowal}, G. and {Yan}, H. and {Beresnyak}, A. and {de Gouveia Dal Pino}, E.~M.},
        title = "{Turbulence, Magnetic Reconnection in Turbulent Fluids and Energetic Particle Acceleration}",
      journal = {Space Science Reviews},
         year = 2012,
        month = nov,
       volume = {173},
       number = {1-4},
        pages = {557-622},
          doi = {10.1007/s11214-012-9936-7},
archivePrefix = {arXiv},
       eprint = {1211.0008},
 primaryClass = {astro-ph.SR},
       adsurl = {https://ui.adsabs.harvard.edu/abs/2012SSRv..173..557L}
}

@article{lesur2021,
  title = {Magnetohydrodynamics of Protoplanetary Discs},
  author = {Lesur, G.},
  year = {2021},
  month = feb,
  journal = {Journal of Plasma Physics},
  volume = {87},
  pages = {205870101},
  issn = {0022-3778},
  doi = {10.1017/S0022377820001002},
  urldate = {2025-02-25}
}

@ARTICLE{Lesur_2023,
       author = {{Lesur}, G.~R.~J. and {Baghdadi}, S. and {Wafflard-Fernandez}, G. and {Mauxion}, J. and {Robert}, C.~M.~T. and {Van den Bossche}, M.},
        title = "{IDEFIX: A versatile performance-portable Godunov code for astrophysical flows}",
      journal = {Journal of Computational Physics},
         year = 2023,
        month = sep,
       volume = {677},
          eid = {A9},
        pages = {A9},
          doi = {10.1051/0004-6361/202346005},
archivePrefix = {arXiv},
       eprint = {2304.13746},
 primaryClass = {astro-ph.IM},
       adsurl = {https://ui.adsabs.harvard.edu/abs/2023A&A...677A...9L}
}

@ARTICLE{liska2022,
       author = {{Liska}, M.~T.~P. and {Chatterjee}, K. and {Issa}, D. and {Yoon}, D. and {Kaaz}, N. and {Tchekhovskoy}, A. and {van Eijnatten}, D. and {Musoke}, G. and {Hesp}, C. and {Rohoza}, V. and {Markoff}, S. and {Ingram}, A. and {van der Klis}, M.},
        title = "{H-AMR: A New GPU-accelerated GRMHD Code for Exascale Computing with 3D Adaptive Mesh Refinement and Local Adaptive Time Stepping}",
      journal = {The Astrophysical Journal Supplement Series},
         year = 2022,
        month = dec,
       volume = {263},
       number = {2},
          eid = {26},
        pages = {26},
          doi = {10.3847/1538-4365/ac9966},
archivePrefix = {arXiv},
       eprint = {1912.10192},
 primaryClass = {astro-ph.HE},
       adsurl = {https://ui.adsabs.harvard.edu/abs/2022ApJS..263...26L}
}

@ARTICLE{Londrillo_DelZanna2004,
   author = {{Londrillo}, P. and {del Zanna}, L.},
    title = "{On the divergence-free condition in Godunov-type schemes for ideal magnetohydrodynamics: the upwind constrained transport method}",
  journal = {Journal of Computational Physics},
   eprint = {astro-ph/0310183},
     year = 2004,
    month = mar,
   volume = 195,
    pages = {17-48},
      doi = {10.1016/j.jcp.2003.09.016},
   adsurl = {http://adsabs.harvard.edu/abs/2004JCoPh.195...17L}
}

@INCOLLECTION{mayer2019,
       author = {{Mayer}, Lucio},
        title = "{Super-Eddington accretion; flow regimes and conditions in high-z galaxies}",
    booktitle = {Formation of the First Black Holes},
         year = 2019,
       editor = {{Latif}, Muhammad and {Schleicher}, Dominik},
        pages = {195-222},
          doi = {10.1142/9789813227958_0011},
       adsurl = {https://ui.adsabs.harvard.edu/abs/2019ffbh.book..195M}
}

@INPROCEEDINGS{mezzacappa2023,
       author = {{Mezzacappa}, Anthony},
        title = "{Toward Realistic Models of Core Collapse Supernovae: A Brief Review}",
    booktitle = {The Predictive Power of Computational Astrophysics as a Discover Tool},
         year = 2023,
       editor = {{Bisikalo}, Dmitry and {Wiebe}, Dmitri and {Boily}, Christian},
       series = {IAU Symposium},
       volume = {362},
        month = jan,
        pages = {215-227},
          doi = {10.1017/S1743921322001831},
archivePrefix = {arXiv},
       eprint = {2205.13438},
 primaryClass = {astro-ph.SR},
       adsurl = {https://ui.adsabs.harvard.edu/abs/2023IAUS..362..215M}
}

@ARTICLE{Mignone2005,
       author = {{Mignone}, A. and {Plewa}, T. and {Bodo}, G.},
        title = "{The Piecewise Parabolic Method for Multidimensional Relativistic Fluid Dynamics}",
      journal = {The Astrophysical Journal Supplement Series},
         year = 2005,
        month = sep,
       volume = {160},
       number = {1},
        pages = {199-219},
          doi = {10.1086/430905},
archivePrefix = {arXiv},
       eprint = {astro-ph/0505200},
 primaryClass = {astro-ph},
       adsurl = {https://ui.adsabs.harvard.edu/abs/2005ApJS..160..199M}
}

@article{Mignone_PLUTO2007,
	Adsurl = {http://adsabs.harvard.edu/abs/2007ApJS..170..228M},
	Author = {{Mignone}, A. and {Bodo}, G. and {Massaglia}, S. and {Matsakos}, T. and {Tesileanu}, O. and {Zanni}, C. and {Ferrari}, A.},
	Doi = {10.1086/513316},
	Eprint = {astro-ph/0701854},
	Journal = {The Astrophysical Journal Supplement Series},
	Month = may,
	Pages = {228-242},
	Title = {{PLUTO: A Numerical Code for Computational Astrophysics}},
	Volume = 170,
	Year = 2007
}

@ARTICLE{Mignone_2010,
   title={High-order conservative finite difference GLM–MHD schemes for cell-centered MHD},
   volume={229},
   ISSN={0021-9991},
   url={http://dx.doi.org/10.1016/j.jcp.2010.04.013},
   DOI={10.1016/j.jcp.2010.04.013},
   number={17},
   journal={Journal of Computational Physics},
   publisher={Elsevier BV},
   author={Mignone, Andrea and Tzeferacos, Petros and Bodo, Gianluigi},
   year={2010},
   month=aug, pages={5896–5920} }

@ARTICLE{Mignone_etal2010,
   author = {{Mignone}, A. and {Rossi}, P. and {Bodo}, G. and {Ferrari}, A. and 
	{Massaglia}, S.},
    title = "{High-resolution 3D relativistic MHD simulations of jets}",
  journal = {Monthly Notices of the Royal Astronomical Society},
archivePrefix = "arXiv",
   eprint = {0908.4523},
     year = 2010,
    month = feb,
   volume = 402,
    pages = {7-12},
      doi = {10.1111/j.1365-2966.2009.15642.x},
   adsurl = {http://adsabs.harvard.edu/abs/2010MNRAS.402....7M}
}

@article{Mignone_PLUTO2012,
	Adsurl = {http://adsabs.harvard.edu/abs/2012ApJS..198....7M},
	Archiveprefix = {arXiv},
	Author = {{Mignone}, A. and {Zanni}, C. and {Tzeferacos}, P. and {van Straalen}, B. and {Colella}, P. and {Bodo}, G.},
	Doi = {10.1088/0067-0049/198/1/7},
	Eid = {7},
	Eprint = {1110.0740},
	Journal = {The Astrophysical Journal Supplement Series},
	Month = jan,
	Pages = {7},
	Primaryclass = {astro-ph.HE},
	Title = {{The PLUTO Code for Adaptive Mesh Computations in Astrophysical Fluid Dynamics}},
	Volume = 198,
	Year = 2012
}

@ARTICLE{Mignone_DelZanna2021,
       author = {{Mignone}, A. and {Del Zanna}, L.},
        title = "{Systematic construction of upwind constrained transport schemes for MHD}",
      journal = {Journal of Computational Physics},
         year = 2021,
        month = jan,
       volume = {424},
          eid = {109748},
        pages = {109748},
          doi = {10.1016/j.jcp.2020.109748},
archivePrefix = {arXiv},
       eprint = {2004.10542},
 primaryClass = {physics.comp-ph},
       adsurl = {https://ui.adsabs.harvard.edu/abs/2021JCoPh.42409748M}
}

@ARTICLE{mignone2024,
       author = {{Mignone}, A. and {Berta}, V. and {Rossazza}, M. and {Bugli}, M. and {Mattia}, G. and {Del Zanna}, L. and {Pareschi}, L.},
        title = "{A fourth-order accurate finite volume scheme for resistive relativistic MHD}",
      journal = {Monthly Notices of the Royal Astronomical Society},
         year = 2024,
        month = sep,
       volume = {533},
       number = {2},
        pages = {1670-1686},
          doi = {10.1093/mnras/stae1729},
archivePrefix = {arXiv},
       eprint = {2407.08519},
 primaryClass = {astro-ph.HE},
       adsurl = {https://ui.adsabs.harvard.edu/abs/2024MNRAS.533.1670M}
}

@ARTICLE{Miyoshi_Kusano2005,
       author = {{Miyoshi}, Takahiro and {Kusano}, Kanya},
        title = "{A multi-state HLL approximate Riemann solver for ideal magnetohydrodynamics}",
      journal = {Journal of Computational Physics},
         year = "2005",
        month = "Sep",
       volume = {208},
       number = {1},
        pages = {315-344},
          doi = {10.1016/j.jcp.2005.02.017},
       adsurl = {https://ui.adsabs.harvard.edu/abs/2005JCoPh.208..315M}
}

@ARTICLE{narayan2005,
       author = {{Narayan}, Ramesh},
        title = "{Black holes in astrophysics}",
      journal = {New Journal of Physics},
         year = 2005,
        month = sep,
       volume = {7},
       number = {1},
        pages = {199},
          doi = {10.1088/1367-2630/7/1/199},
archivePrefix = {arXiv},
       eprint = {gr-qc/0506078},
 primaryClass = {gr-qc},
       adsurl = {https://ui.adsabs.harvard.edu/abs/2005NJPh....7..199N}
}

@ARTICLE{2021_Lesur,
       author = {{Lesur}, Geoffroy R.~J.},
        title = "{Systematic description of wind-driven protoplanetary discs}",
      journal = {Journal of Computational Physics},
         year = 2021,
        month = jun,
       volume = {650},
          eid = {A35},
        pages = {A35},
          doi = {10.1051/0004-6361/202040109},
archivePrefix = {arXiv},
       eprint = {2101.10349},
 primaryClass = {astro-ph.SR},
       adsurl = {https://ui.adsabs.harvard.edu/abs/2021A&A...650A..35L}
}

@ARTICLE{bugli2025,
       author = {{Bugli}, M. and {Lopresti}, E.~F. and {Figueiredo}, E. and {Mignone}, A. and {Cerutti}, B. and {Mattia}, G. and {Del Zanna}, L. and {Bodo}, G. and {Berta}, V.},
        title = "{Relativistic reconnection with effective resistivity: I. Dynamics and reconnection rate}",
      journal = {Journal of Computational Physics},
         year = 2025,
        month = jan,
       volume = {693},
          eid = {A233},
        pages = {A233},
          doi = {10.1051/0004-6361/202452277},
archivePrefix = {arXiv},
       eprint = {2410.20924},
 primaryClass = {astro-ph.HE},
       adsurl = {https://ui.adsabs.harvard.edu/abs/2025A&A...693A.233B}
}

@ARTICLE{2019_Olmi,
       author = {{Olmi}, B. and {Bucciantini}, N.},
        title = "{On the origin of jet-like features in bow shock pulsar wind nebulae}",
      journal = {Monthly Notices of the Royal Astronomical Society},
         year = 2019,
        month = dec,
       volume = {490},
       number = {3},
        pages = {3608-3615},
          doi = {10.1093/mnras/stz2819},
archivePrefix = {arXiv},
       eprint = {1910.01926},
 primaryClass = {astro-ph.HE},
       adsurl = {https://ui.adsabs.harvard.edu/abs/2019MNRAS.490.3608O}
}

@ARTICLE{2018_Olmi,
       author = {{Olmi}, B. and {Bucciantini}, N. and {Morlino}, G.},
        title = "{Numerical simulations of mass loading in the tails of bow-shock pulsar-wind nebulae}",
      journal = {Monthly Notices of the Royal Astronomical Society},
         year = 2018,
        month = dec,
       volume = {481},
       number = {3},
        pages = {3394-3400},
          doi = {10.1093/mnras/sty2525},
archivePrefix = {arXiv},
       eprint = {1809.03807},
 primaryClass = {astro-ph.HE},
       adsurl = {https://ui.adsabs.harvard.edu/abs/2018MNRAS.481.3394O}
}

@ARTICLE{orlando2011,
       author = {{Orlando}, Salvatore and {Reale}, Fabio and {Peres}, Giovanni and {Mignone}, Andrea},
        title = "{Mass accretion to young stars triggered by flaring activity in circumstellar discs}",
      journal = {Monthly Notices of the Royal Astronomical Society},
         year = 2011,
        month = aug,
       volume = {415},
       number = {4},
        pages = {3380-3392},
          doi = {10.1111/j.1365-2966.2011.18954.x},
archivePrefix = {arXiv},
       eprint = {1104.5107},
 primaryClass = {astro-ph.SR},
       adsurl = {https://ui.adsabs.harvard.edu/abs/2011MNRAS.415.3380O}
}

@ARTICLE{Powell_etal1999,
       author = {{Powell}, Kenneth G. and {Roe}, Philip L. and {Linde}, Timur J. and
         {Gombosi}, Tamas I. and {De Zeeuw}, Darren L.},
        title = "{A Solution-Adaptive Upwind Scheme for Ideal Magnetohydrodynamics}",
      journal = {Journal of Computational Physics},
         year = "1999",
        month = "Sep",
       volume = {154},
       number = {2},
        pages = {284-309},
          doi = {10.1006/jcph.1999.6299},
       adsurl = {https://ui.adsabs.harvard.edu/abs/1999JCoPh.154..284P}
}

@article{mignoneaoptimization,
  title={I/O Optimization Strategies in the PLUTO Code},
  author={Mignone, A and Muscianisib, G and Rivib, M and Bodoc, G}
}

@ARTICLE{Mignone_2013,
       author = {{Mignone}, A. and {Striani}, E. and {Tavani}, M. and {Ferrari}, A.},
        title = "{Modelling the kinked jet of the Crab nebula}",
      journal = {Monthly Notices of the Royal Astronomical Society},
         year = 2013,
        month = dec,
       volume = {436},
       number = {2},
        pages = {1102-1115},
          doi = {10.1093/mnras/stt1632},
archivePrefix = {arXiv},
       eprint = {1309.0375},
 primaryClass = {astro-ph.HE},
       adsurl = {https://ui.adsabs.harvard.edu/abs/2013MNRAS.436.1102M}
}

@ARTICLE{Pavan_2023,
       author = {{Pavan}, Andrea and {Ciolfi}, Riccardo and {Kalinani}, Jay V. and {Mignone}, Andrea},
        title = "{Jet-environment interplay in magnetized binary neutron star mergers}",
      journal = {Monthly Notices of the Royal Astronomical Society},
         year = 2023,
        month = sep,
       volume = {524},
       number = {1},
        pages = {260-275},
          doi = {10.1093/mnras/stad1809},
archivePrefix = {arXiv},
       eprint = {2211.10135},
 primaryClass = {astro-ph.HE},
       adsurl = {https://ui.adsabs.harvard.edu/abs/2023MNRAS.524..260P}
}

@article{perri2018,
  title = {Simulations of Solar Wind Variations during an 11-Year Cycle and the Influence of North-South Asymmetry},
  author = {Perri, B. and Brun, A. S. and R{\'e}ville, V. and Strugarek, A.},
  year = {2018},
  month = oct,
  journal = {Journal of Plasma Physics},
  volume = {84},
  pages = {765840501},
  issn = {0022-3778},
  doi = {10.1017/S0022377818000880},
  urldate = {2025-02-28}
}

@ARTICLE{2022_Bodo,
       author = {{Bodo}, G. and {Mamatsashvili}, G. and {Rossi}, P. and {Mignone}, A.},
        title = "{Current-driven kink instabilities in relativistic jets: dissipation properties}",
      journal = {Monthly Notices of the Royal Astronomical Society},
         year = 2022,
        month = feb,
       volume = {510},
       number = {2},
        pages = {2391-2406},
          doi = {10.1093/mnras/stab3492},
archivePrefix = {arXiv},
       eprint = {2111.14575},
 primaryClass = {astro-ph.HE},
       adsurl = {https://ui.adsabs.harvard.edu/abs/2022MNRAS.510.2391B}
}

@ARTICLE{2021_Flock,
       author = {{Flock}, Mario and {Mignone}, Andrea},
        title = "{Streaming instability in a global patch simulation of protoplanetary disks}",
      journal = {Journal of Computational Physics},
         year = 2021,
        month = jun,
       volume = {650},
          eid = {A119},
        pages = {A119},
          doi = {10.1051/0004-6361/202040104},
archivePrefix = {arXiv},
       eprint = {2103.15146},
 primaryClass = {astro-ph.EP},
       adsurl = {https://ui.adsabs.harvard.edu/abs/2021A&A...650A.119F}
}

@ARTICLE{2020_Orlando,
       author = {{Orlando}, S. and {Ono}, M. and {Nagataki}, S. and {Miceli}, M. and {Umeda}, H. and {Ferrand}, G. and {Bocchino}, F. and {Petruk}, O. and {Peres}, G. and {Takahashi}, K. and {Yoshida}, T.},
        title = "{Hydrodynamic simulations unravel the progenitor-supernova-remnant connection in SN 1987A}",
      journal = {Journal of Computational Physics},
         year = 2020,
        month = apr,
       volume = {636},
          eid = {A22},
        pages = {A22},
          doi = {10.1051/0004-6361/201936718},
archivePrefix = {arXiv},
       eprint = {1912.03070},
 primaryClass = {astro-ph.HE},
       adsurl = {https://ui.adsabs.harvard.edu/abs/2020A&A...636A..22O}
}

@ARTICLE{2019_Orlando,
       author = {{Orlando}, S. and {Miceli}, M. and {Petruk}, O. and {Ono}, M. and {Nagataki}, S. and {Aloy}, M.~A. and {Mimica}, P. and {Lee}, S. -H. and {Bocchino}, F. and {Peres}, G. and {Guarrasi}, M.},
        title = "{3D MHD modeling of the expanding remnant of SN 1987A. Role of magnetic field and non-thermal radio emission}",
      journal = {Journal of Computational Physics},
         year = 2019,
        month = feb,
       volume = {622},
          eid = {A73},
        pages = {A73},
          doi = {10.1051/0004-6361/201834487},
archivePrefix = {arXiv},
       eprint = {1812.00021},
 primaryClass = {astro-ph.HE},
       adsurl = {https://ui.adsabs.harvard.edu/abs/2019A&A...622A..73O}
}

@ARTICLE{Mattia_2023,
       author = {{Mattia}, G. and {Del Zanna}, L. and {Bugli}, M. and {Pavan}, A. and {Ciolfi}, R. and {Bodo}, G. and {Mignone}, A.},
        title = "{Resistive relativistic MHD simulations of astrophysical jets}",
      journal = {Journal of Computational Physics},
         year = 2023,
        month = nov,
       volume = {679},
          eid = {A49},
        pages = {A49},
          doi = {10.1051/0004-6361/202347126},
archivePrefix = {arXiv},
       eprint = {2308.09477},
 primaryClass = {astro-ph.HE},
       adsurl = {https://ui.adsabs.harvard.edu/abs/2023A&A...679A..49M}
}

@ARTICLE{2010_Mignone,
       author = {{Mignone}, A. and {Rossi}, P. and {Bodo}, G. and {Ferrari}, A. and {Massaglia}, S.},
        title = "{High-resolution 3D relativistic MHD simulations of jets}",
      journal = {Monthly Notices of the Royal Astronomical Society},
         year = 2010,
        month = feb,
       volume = {402},
       number = {1},
        pages = {7-12},
          doi = {10.1111/j.1365-2966.2009.15642.x},
archivePrefix = {arXiv},
       eprint = {0908.4523},
 primaryClass = {astro-ph.CO},
       adsurl = {https://ui.adsabs.harvard.edu/abs/2010MNRAS.402....7M}
}

@article{reville2015,
  title = {The {{Effect}} of {{Magnetic Topology}} on {{Thermally Driven Wind}}: {{Toward}} a {{General Formulation}} of the {{Braking Law}}},
  shorttitle = {The {{Effect}} of {{Magnetic Topology}} on {{Thermally Driven Wind}}},
  author = {R{\'e}ville, Victor and Brun, Allan Sacha and Matt, Sean P. and Strugarek, Antoine and Pinto, Rui F.},
  year = {2015},
  month = jan,
  journal = {The Astrophysical Journal},
  volume = {798},
  pages = {116},
  publisher = {IOP},
  issn = {0004-637X},
  doi = {10.1088/0004-637X/798/2/116},
  urldate = {2025-02-28}
}

@ARTICLE{shibata2011,
       author = {{Shibata}, Kazunari and {Magara}, Tetsuya},
        title = "{Solar Flares: Magnetohydrodynamic Processes}",
      journal = {Living Reviews in Solar Physics},
         year = 2011,
        month = dec,
       volume = {8},
       number = {1},
          eid = {6},
        pages = {6},
          doi = {10.12942/lrsp-2011-6},
       adsurl = {https://ui.adsabs.harvard.edu/abs/2011LRSP....8....6S}
}

@ARTICLE{ShuOsher1988,
       author = {{Shu}, Chi-Wang and {Osher}, Stanley},
        title = "{Efficient Implementation of Essentially Non-oscillatory Shock-Capturing Schemes}",
      journal = {Journal of Computational Physics},
         year = 1988,
        month = aug,
       volume = {77},
       number = {2},
        pages = {439-471},
          doi = {10.1016/0021-9991(88)90177-5},
       adsurl = {https://ui.adsabs.harvard.edu/abs/1988JCoPh..77..439S}
}

@ARTICLE{stone2024,
       author = {{Stone}, James M. and {Mullen}, Patrick D. and {Fielding}, Drummond and {Grete}, Philipp and {Guo}, Minghao and {Kempski}, Philipp and {Most}, Elias R. and {White}, Christopher J. and {Wong}, George N.},
        title = "{AthenaK: A Performance-Portable Version of the Athena++ AMR Framework}",
      journal = {arXiv e-prints},
         year = 2024,
        month = sep,
          eid = {arXiv:2409.16053},
        pages = {arXiv:2409.16053},
          doi = {10.48550/arXiv.2409.16053},
archivePrefix = {arXiv},
       eprint = {2409.16053},
 primaryClass = {astro-ph.IM},
       adsurl = {https://ui.adsabs.harvard.edu/abs/2024arXiv240916053S}
}

@ARTICLE{Suresh_Huynh1997,
   author = {{Suresh}, A. and {Huynh}, H.~T.},
    title = "{Accurate Monotonicity-Preserving Schemes with Runge Kutta Time Stepping}",
  journal = {Journal of Computational Physics},
     year = 1997,
    month = sep,
   volume = 136,
    pages = {83-99},
      doi = {10.1006/jcph.1997.5745},
   adsurl = {http://adsabs.harvard.edu/abs/1997JCoPh.136...83S}
}

@article{teyssier2019,
  title = {Numerical {{Methods}} for {{Simulating Star Formation}}},
  author = {Teyssier, Romain and Commer{\c c}on, Beno{\^i}t},
  year = {2019},
  month = jul,
  journal = {Frontiers in Astronomy and Space Sciences},
  volume = {6},
  pages = {51},
  doi = {10.3389/fspas.2019.00051},
  urldate = {2025-02-25}
}

@ARTICLE{uzdensky2014,
       author = {{Uzdensky}, Dmitri A. and {Rightley}, Shane},
        title = "{Plasma physics of extreme astrophysical environments}",
      journal = {Reports on Progress in Physics},
         year = 2014,
        month = mar,
       volume = {77},
       number = {3},
          eid = {036902},
        pages = {036902},
          doi = {10.1088/0034-4885/77/3/036902},
archivePrefix = {arXiv},
       eprint = {1401.5110},
 primaryClass = {astro-ph.HE},
       adsurl = {https://ui.adsabs.harvard.edu/abs/2014RPPh...77c6902U}
}

@article{zhang2018,
  title = {A {{Review}} of the {{Theory}} of {{Galactic Winds Driven}} by {{Stellar Feedback}}},
  author = {Zhang, Dong},
  year = {2018},
  month = nov,
  journal = {Galaxies},
  volume = {6},
  pages = {114},
  doi = {10.3390/galaxies6040114},
  urldate = {2025-02-25}
}
\end{document}